\documentclass[11pt,english]{article}
\usepackage[T1]{fontenc}
\usepackage[latin1]{inputenc}
\usepackage{a4wide}
\usepackage{amsmath}
\usepackage{graphicx}
\usepackage{amssymb}

\makeatletter
 \newcommand{\lyxaddress}[1]{
   \par {\raggedright #1 
   \vspace{1.4em}
   \noindent\par}
 }

\makeatletter 

\renewcommand\theequation{\hbox{\normalsize\arabic{section}.\arabic{equation}}} \@addtoreset{equation}{section}

\renewcommand\thefigure{\hbox{\normalsize\arabic{section}.\arabic{figure}}} \@addtoreset{figure}{section}

\renewcommand\thetable{\hbox{\normalsize\arabic{section}.\arabic{table}}} \@addtoreset{table}{section}

\makeatother

\usepackage{babel}
\makeatother
\begin{document}

\title{Boundary one-point function, Casimir energy and boundary state formalism
in $D+1$ dimensional QFT}

\author{Z. Bajnok$^{1}$, L. Palla$^{2}$, and G. Takács$^{1}$}

\maketitle

\lyxaddress{\begin{center}$^{1}$\emph{\small Theoretical Physics Research Group,
Hungarian Academy of Sciences, }\\
\emph{\small 1117 Budapest, Pázmány Péter sétány 1/A, Hungary} \\
$^{2}$\emph{\small Institute for Theoretical Physics, Eötvös University,
}\\
\emph{\small 1117 Budapest, Pázmány Péter sétány 1/A, Hungary}\end{center}}

\begin{abstract}
We consider quantum field theories with boundary on a codimension
one hyperplane. Using $1+1$ dimensional examples, we clarify the
relation between three parameters characterising one-point functions,
finite size corrections to the ground state energy and the singularity
structure of scattering amplitudes, respectively. We then develop
the formalism of boundary states in general $D+1$ spacetime dimensions
and relate the cluster expansion of the boundary state to the correlation
functions using reduction formulae. This allows us to derive the cluster
expansion in terms of the boundary scattering amplitudes, and to give
a derivation of the conjectured relations between the parameters in
$1+1$ dimensions, and their generalization to $D+1$ dimensions.
We use these results to express the large volume asymptotics of the
Casimir effect in terms of the one-point functions or alternatively
the singularity structure of the one-particle reflection factor, and
for the case of vanishing one-particle couplings we give a complete
proof of our previous result for the leading behaviour.
\end{abstract}

\section{Introduction}

In this paper we treat quantum field theories with a boundary, which
is supposed to be a hyperplane (of codimension $1$) in flat spacetime.
We also treat finite size effects when two such boundaries are parallel
to each other and separated by a distance $L$ (a.k.a. the Casimir
effect in a planar situation).

Such theories have been extensively studied in $1+1$ dimensions since
the seminal paper of Ghoshal and Zamolodchikov \cite{gz}, which introduced
some fundamental ideas into the subject. They defined and characterized
the so-called boundary state in integrable field theories%
\footnote{In Section 3 we give a definition for the boundary state in any boundary
quantum field theory in arbitrary number of spacetime dimensions,
but the details are not necessary for the time being.%
}, which makes possible the description of boundary phenomena using
the Hamiltonian formalism of the bulk theory. Using this concept,
they set up the necessary notions to study integrable boundary QFTs
by extending the ideas of analytic $S$ matrix theory, factorization
and bootstrap already well-known in the bulk situation. 

Boundary field theories in $1+1$ dimensions are relevant to numerous
condensed matter phenomena (the most prominent examples are the so-called
{}``impurity'' problems), to nonperturbative aspects of string theory
({}``branes'') and also are of interest from the theoretical point
of view as a tool to understand quantum field theory from a different
perspective. However, boundary problems are also relevant in higher
dimensional quantum field theories as descriptions of surface critical
phenomena, and, as mentioned above, in the context of the Casimir
effect.

One of the very interesting phenomena that can occur in the presence
of a boundary is the existence of nontrivial ground state configurations,
which approach the bulk vacuum only asymptotically, far away from
the boundary. In a theory with a mass gap $m$ in the bulk%
\footnote{In the following we always suppose that this is the case, i.e. we
consider only theories with a non-vanishing mass gap, for which the
formalism of asymptotic states is well defined.%
}, any vacuum expectation value (VEV) of a field $\phi$ is bound to
approach the bulk value $\phi_{0}$ exponentially\begin{equation}
\langle\phi(t,\underbar{x})\rangle_{\alpha}\sim\phi_{0}+\bar{g}_{\alpha}\,\mathrm{e}^{-md}\label{eq:gbar_asymptotics}\end{equation}
as a function of the distance $d$ from the boundary, where $\bar{g}_{\alpha}$
is a quantity characterising the boundary condition labelled by $\alpha$.
We mention that $\bar{g}$ itself is a physically relevant quantity
in certain applications, e.g. in the context of thermal Coulomb plasma
in contact with an ideal conductor electrode where it corresponds
to the screened (or 'renormalized') surface charge density \cite{samaj_bajnok}.
The existence of the $\bar{g}$ term in (\ref{eq:gbar_asymptotics})
can be related to the presence of a one-particle term in the boundary
state of the form \cite{dptw_onepoint}\begin{equation}
|B_{\alpha}\rangle=(1+\tilde{g}_{\alpha}A^{+}(\underbar{0})+\dots)|0\rangle\label{eq:bstate_onepart}\end{equation}
where $A^{+}\left(\underbar{k}\right)$ - with the normalization $[A(\underbar{k}),\, A^{+}(\underbar{k}^{\prime})]=(2\pi)^{D}\sqrt{\underbar{k}^{2}+m^{2}}\delta(\underbar{k}-\underbar{k}^{\prime})$
- denotes the operator that creates an asymptotic particle of momentum
$\underbar{k}$, and $\tilde{g}_{\alpha}$ is just another quantity
characterising the strength of the one-particle contribution to the
boundary state. It is straightforward to derive (see subsection 3.2)
the relation \begin{equation}
\bar{g}_{\alpha}=\sqrt{\frac{Z}{2}}\tilde{g}_{\alpha}\label{eq:gbar_gtilde_intro}\end{equation}
where $Z$ is the wavefunction renormalization constant of $\phi$,
considered as an interpolating field for the bulk asymptotic multi-particle
states:

\[
\langle0|\phi(0)|A(\underbar{k}=\underbar{0})\rangle=\sqrt{\frac{Z}{2}}\]
where $|0\rangle$ is the bulk vacuum state and $|A(\underbar{k}=\underbar{0})\rangle$
is the asymptotic one-particle state containing a single particle
with the lowest mass $m$ and momentum zero. Note that $Z$ is a bulk
quantity which is independent of the boundary condition.

It was also shown in \cite{Lusch} that for $1+1$ dimensional QFTs
the presence of a one-particle term in the boundary state leads to
the following asymptotics of the ground state (Casimir) energy on
a strip with boundary conditions $\alpha$ and $\beta$\begin{equation}
E_{\alpha\beta}(L)\sim-m\tilde{g}_{\alpha}\tilde{g}_{\beta}\mathrm{e}^{-mL}\label{eq:Casimir_onepart}\end{equation}
while in the absence of such one-particle coupling the energy is expected
to decay as $\mathrm{e}^{-2mL}$ (which was already known for the
integrable case from studies of the thermodynamical Bethe Ansatz \cite{lmss}).
In subsection 3.5 we show that this is also valid in general $D+1$
spacetime dimensions, and provides a more convenient way of defining
the quantity $\tilde{g}_{\alpha}$ than the expansion (\ref{eq:bstate_onepart}).
The definition of $\tilde{g}_{\alpha}$ from the one-particle term
in the boundary state depends on the convention of normalising the
one-particle state, while the Casimir energy (\ref{eq:Casimir_onepart})
is a directly measurable quantity independent of any conventions in
the field theoretic formalism.

It was already noted in \cite{gz} that there is another manifestation
of a one-particle coupling to the boundary, namely that the one-particle
reflection factor has a pole at a special location. Introducing the
rapidity parametrization of the $1+1$ dimensional energy-momentum
$(e,\, p)$ as usual\[
e=m\cosh\theta\quad,\quad p=m\sinh\theta\]
it can be argued that the one-particle reflection factor off the boundary
(which is the amplitude for the process involving a single particle
both in the initial and in the final state) must have a pole at $\theta=i\frac{\pi}{2}$
with the residue denoted by

\begin{equation}
\mathop{\mathrm{Res}}_{\theta=i\frac{\pi}{2}}R_{\alpha}\left(\theta\right)=i\frac{g_{\alpha}^{2}}{2}\label{eq:pi2pole}\end{equation}
which defines another characteristic quantity $g_{\alpha}$ (which
is real for unitary theories)%
\footnote{Note that the above $g$ quantities have nothing to do with the boundary
entropy introduced in \cite{affleck_ludwig} which is also usually
denoted by $g$ and is often called the $g$-function.%
}. Ghoshal and Zamolodchikov identified $g_{\alpha}$ with $\tilde{g}_{\alpha}$,
but later Dorey et al. \cite{dptw_onepoint}, investigating one-point
functions in the so-called scaling Lee-Yang model found numerically
the relation\begin{equation}
\tilde{g}_{\alpha}=\frac{g_{\alpha}}{2}\label{eq:gtilde_g_intro}\end{equation}
In the previous papers \cite{Lusch,stripcikk,bbf} we presented evidence
(both analytic and numerical) that this relation extends to all $1+1$
dimensional integrable quantum field theories. Strictly speaking,
however, there existed no field theoretic derivation or proof of this
relation up to now. It was not clear either whether this relation
can be extended to general boundary QFTs. 

The knowledge of the relation between the $g$ parameters has another
interesting application. Namely, the classical vacuum expectation
value and so the classical limit of $\bar{g}_{\alpha}$ can be calculated
explicitely by solving the classical field equations. Relating this
quantity to the residue of the reflection factor ($g_{\alpha}$) makes
it possible to identify the boundary condition corresponding to reflection
factors found by solving the boundary bootstrap conditions, which
is the fundamental idea underlying the work by Fateev and Onofri in
\cite{fateev_onofri}. 

The central aim of the present work is to extend these results already
known in $1+1$ dimensional integrable QFTs, to nonintegrable theories
and further to boundary QFTs defined in any spacetime dimensions.
Even in the case of integrable theories, the derivation of (\ref{eq:gtilde_g_intro})
presented here is the first one that is truly general and starts from
{}``first principles'' (i.e. does not depend on additional assumptions
or approximations).

We start in section 2 by reviewing the status in $1+1$ dimensional
QFTs. We present arguments in the integrable case, further to those
already made in \cite{Lusch}, and then extend our considerations
to the nonintegrable case using semiclassical techniques, where we
show that both (\ref{eq:gbar_gtilde_intro}) and (\ref{eq:gtilde_g_intro})
can be extended to this case. However, there is no way to tackle general
nonintegrable quantum field theories without further theoretical developments.

In section 3, therefore, we develop the necessary tools to address
the problem, using only general concepts of quantum field theory which
are valid in any spacetime dimensions. In order to develop the boundary
state formalism, in the appendix we derive a set of reduction formulae
relating the matrix elements of the boundary state with bulk asymptotic
many-particle states to correlation functions. Using them we can describe
the boundary state in terms of (analytic continuation of) the boundary
scattering amplitudes, analogously to what Ghoshal and Zamolodchikov
did in the case of $1+1$ dimensional integrable field theories \cite{gz}.
We present the detailed derivation of the one-particle term in the
boundary state which leads us to the relation (\ref{eq:gbar_gtilde_intro}),
and of the two-particle term, which is given in terms of the one-particle
reflection factor. We then use a cluster argument to relate the singularity
of the reflection factor to the one-point function, i.e. $g$ to $\bar{g}$.
For the case of $1+1$ dimensions this gives a proof of relation (\ref{eq:gtilde_g_intro})
{}``from first principles'', valid for integrable and nonintegrable
QFTs as well, while in the case of $D+1$ dimensional theories it
allows us to characterize the nature and strength of the appropriate
singularity of the reflection factor. In subsection 3.5 we give the
derivation underlying the claim in our previous paper \cite{Casimir}
where we gave a universal expression for the leading asymptotics of
the Casimir energy (in the planar situation) valid also for nontrivially
interacting theories, and also show how these results are affected
by the presence of one-particle couplings to the boundary. We draw
our final conclusions in section 4.

\section{Examples}

In this section we consider several examples as well as some heuristic
derivations (under various assumptions) of the conjectured relations
among the various $g$-s. Some of the examples and derivations are
worked out in specific models, while some of the considerations are
model independent. Part of the material in this section is on the
level of (semi) classical considerations while the other is on the
level of full QFT. All the examples are worked out in $1+1$ dimensions;
they intended to provide a clear understanding of the underlying physics,
and motivation for the derivations presented in the section 3, which
are model independent and valid for generic number of spacetime dimensions.

\subsection{Quantum theory examples}

\subsubsection{Boundary sine-Gordon model with Dirichlet boundary conditions}

Our first example is the boundary sine-Gordon model with Dirichlet
boundary conditions. In this model the bulk Lagrangian is written
as \begin{equation}
\mathcal{L}_{SG}(x,t)=\frac{1}{2}\left(\partial_{\mu}\Phi\right)^{2}-\frac{m^{2}}{\beta^{2}}\left(1-\cos\beta\Phi\right)\label{eq:sgbulk}\end{equation}
 where $\beta$ is the bulk coupling, $m$ is the mass parameter and
the field satisfies the boundary conditions \[
\Phi(x,t)|_{x=0}=\Phi_{0}^{D}\,,\qquad\Phi(x,t)|_{x=L}=\Phi_{L}^{D}\]
 The fundamental range of the parameters $\Phi_{0,L}^{D}$ is given
by $0\leq\frac{\beta}{2}\Phi_{0,L}^{D}\equiv\varphi_{0,L}\leq\frac{\pi}{2}$.
In the regime $\beta^{2}<4\pi$ the lightest particle in the spectrum
is the first breather, and it has the following reflection amplitude
off the boundary \cite{ghoshal}\begin{equation}
R_{|\rangle}^{(1)}(\theta)=\frac{\left(\frac{1}{2}\right)\left(\frac{1}{2\lambda}+1\right)}{\left(\frac{1}{2\lambda}+\frac{3}{2}\right)}\frac{\left(\frac{\eta}{\pi\lambda}-\frac{1}{2}\right)}{\left(\frac{\eta}{\pi\lambda}+\frac{1}{2}\right)}\;,\quad\quad(x)=\frac{\sinh\left(\frac{\theta}{2}+i\frac{\pi x}{2}\right)}{\sinh\left(\frac{\theta}{2}-i\frac{\pi x}{2}\right)}\label{eq:b1refl}\end{equation}
 Here $\theta$ is the rapidity of $B^{1}$ while \[
\lambda=\frac{8\pi}{\beta^{2}}-1\,,\qquad{\textrm{and}}\qquad\eta_{0,L}=\mp(\lambda+1)\varphi_{0,L}\]
 are the parameters characterising the bootstrap solution \cite{gz}.
Note that $R_{|\rangle}^{(1)}(\theta)$ has a first order pole at
$\theta=i\frac{\pi}{2}$ originating from the $\left(\frac{1}{2}\right)$
factor. This results in the following coupling strength between the
first breather and the boundary:\begin{equation}
g_{1}\left(\eta\right)=2\sqrt{\frac{1+\cos\frac{\pi}{2\lambda}-\sin\frac{\pi}{2\lambda}}{1-\cos\frac{\pi}{2\lambda}+\sin\frac{\pi}{2\lambda}}}\tan\frac{\eta}{2\lambda}\label{eq:sgg1}\end{equation}
 (The expression under the square root is always positive as long
as $\lambda>1$ which is necessary for the first breather to exist
in the spectrum). However, some care must be taken, because the sign
of the coupling $g_{1}$ must be opposite on the two ends of the strip
since these are related by a spatial reflection under which the first
breather is odd. As a result, eqn. (\ref{eq:Casimir_onepart}) and
eqn. (\ref{eq:gtilde_g_intro}) predict that the leading finite size
correction to the ground state energy on the strip is given by \begin{equation}
E(\varphi_{0},\varphi_{L})=m_{1}\frac{g_{1}(\eta_{0})g_{1}(\eta_{L})}{4}e^{-mL}+\dots\label{eq:b1ener}\end{equation}
 In \cite{Lusch} this expression was checked by comparing it to the
exact ground state energy at large but finite $L$-s. The exact ground
state energy of the boundary sine-Gordon model with Dirichlet boundary
conditions was computed numerically (to high precision) from the NLIE
proposed recently \cite{abr} and the detailed numerical comparison
showed an excellent agreement. This comparison was later extended
to more general boundary conditions in \cite{ahn_et_al}.

\subsubsection{BTBA in the infrared limit}

In this section we calculate the infrared limit of boundary thermodynamic
Bethe Ansatz (BTBA) to support the relation (\ref{eq:gtilde_g_intro}).
Although the original derivation of BTBA given by Leclair et al. \cite{lmss}
is only valid for the case when there is no one-particle term in the
boundary state, it can be easily argued that the presence of one-particle
terms makes no difference to the end result, and this is also supported
by numerical studies using comparison with truncated conformal space
(TCS) by Dorey et al. in \cite{dptw_btba}.

In a theory with a single particle of mass $m$ on a strip of length
$L$ the BTBA equation for the pseudo energy $\epsilon(\theta)$ takes
the form\begin{equation}
\epsilon\left(\theta\right)=2mL\cosh\theta-\int_{-\infty}^{\infty}\frac{d\theta'}{2\pi}\Phi\left(\theta-\theta'\right)\log\left(1+\chi\left(\theta'\right)\mathrm{e}^{-\epsilon(\theta')}\right)\label{eq:BTBA_eqn}\end{equation}
 where $\Phi\left(\theta\right)=-i\frac{d}{d\theta}\log S\left(\theta\right)$
is the derivative of the phase of the two-particle $S$-matrix $S(\theta)$,
and $\chi\left(\theta\right)=\bar{K}_{\alpha}\left(\theta\right)K_{\beta}\left(\theta\right)$
where \[
K_{\alpha}(\theta)=R_{\alpha}\left(i\frac{\pi}{2}-\theta\right)\]
and $\bar{K}_{\alpha}(\theta)=K_{\alpha}(-\theta)$. Once $\epsilon(\theta)$
is obtained, the energy of the ground state is given by \[
E_{\alpha\beta}(L)=-m\int_{-\infty}^{\infty}\frac{d\theta}{4\pi}\cosh\theta\log\left(1+\chi\left(\theta\right)\mathrm{e}^{-\epsilon(\theta)}\right)\]
 In a theory with nonzero one-particle coupling on both boundaries
$\chi$ has a second-order pole at $\theta=0$. The logarithmic terms
in (\ref{eq:BTBA_eqn}) and in the ground state energy remain integrable
and the BTBA makes perfectly good sense, but to obtain the correct
asymptotic ($L\rightarrow\infty$) expression one needs to be careful.
For large $L$ we finally obtained in \cite{Lusch}\begin{equation}
E_{\alpha\beta}(L)=-m\frac{\left|g_{\alpha}g_{\beta}\right|}{4}\mathrm{e}^{-mL}-m\int\limits _{-\infty}^{\infty}\frac{d\theta}{4\pi}\cosh\theta\log\left(\frac{1+\chi\left(\theta\right)\mathrm{e}^{-2mL\cosh\theta}}{1+\frac{g_{\alpha}^{2}g_{\beta}^{2}}{4\sinh^{2}\theta}\mathrm{e}^{-2mL}}\right)\label{eq:finalener}\end{equation}
 The remaining integral is an expression of order $\mathrm{e}^{-2mL}$,
while the leading term agrees with (\ref{eq:Casimir_onepart}) (for
$\tilde{g}_{\beta}=g_{\beta}/2$) if $g_{\alpha}g_{\beta}>0$. It
was already noted in \cite{dptw_btba} (using a comparison with truncated
conformal space method) that the BTBA equation only gives the correct
ground state energy in this case.

Here we extend this result to values of boundary parameters such that
$g_{\alpha}g_{\beta}<0$, applying a suitable analytic continuation.
Note that in the definition of the boundary coupling $g_{\beta}$
a branch of the square root function must be chosen. In all known
cases (e.g. Lee-Yang in \cite{dptw_btba} or sine-Gordon) there exists
a branch choice such that the boundary couplings depend analytically
on the boundary parameters. In this case a straightforward analytic
continuation of the TBA equations (following the ideas originally
proposed in \cite{dorey-tateo}) from the domain where $g_{\alpha}g_{\beta}>0$
to the one with $g_{\alpha}g_{\beta}<0$ gives the ground state energy
as \[
E_{\alpha\beta}(L)=m\sin u-m\int_{-\infty}^{\infty}\frac{d\theta}{4\pi}\cosh\theta\log\left(1+\chi\left(\theta\right)\mathrm{e}^{-\epsilon(\theta)}\right)\]
 where now $\epsilon(\theta)$ is the solution of the equation \begin{equation}
\epsilon\left(\theta\right)=2mL\cosh\theta-\log\frac{S(\theta-iu)}{S(\theta+iu)}+\int_{-\infty}^{\infty}\frac{d\theta'}{2\pi}\Phi\left(\theta-\theta'\right)\log\left(1+\chi\left(\theta'\right)\mathrm{e}^{-\epsilon(\theta')}\right)\label{eq:BTBA_eqnf}\end{equation}
 and $u$ is determined by the equation \[
\bar{K}_{\beta}(iu)K_{\alpha}(iu)e^{-\epsilon(iu)}=-1\]
 For $L$ large one can put $\epsilon(\theta)=2mL\cosh\theta$, yielding
$\sin u\sim u\sim\frac{|g_{\alpha}g_{\beta}|}{2}e^{-mL}$, which in
the energy expression just flips the sign of the term coming from
the integral part. Thus \[
E_{\alpha\beta}(L)=-m\frac{g_{\alpha}g_{\beta}}{4}\mathrm{e}^{-mL}+\dots\]
 in the parameter range where $g_{\alpha}g_{\beta}<0$, which is consistent
with eqns (\ref{eq:gtilde_g_intro},\ref{eq:Casimir_onepart}) which
is the result of the cluster expansion \cite{Lusch}.

\subsubsection{Bethe-Yang equation}

The aim of the following considerations is to emphasize the difference
between the poles of the reflection amplitude corresponding to boundary
bound states and the one appearing at $\theta=i\pi/2$, and to give
an alternative argument for the validity of eqn. (\ref{eq:gtilde_g_intro}).

Suppose that in the half line theory there is a boundary bound state
which is described by a pole of the reflection amplitude \[
R_{\alpha}(\theta)\sim\frac{i\Gamma^{2}/2}{\theta-iv_{0}}\]
 in the physical strip $0<v_{0}<\frac{\pi}{2}$. The simplest way
to determine the leading finite size correction to the energy of the
bound state is to confine the theory to a strip of width $L$ by imposing
identical boundary conditions at the ends of the strip and look for
purely imaginary solutions of the Bethe-Yang equation of the fundamental
particle%
\footnote{This equation neglects vacuum polarization corrections, but one can
argue that for large $L$-s it correctly gives the leading contribution
- this was checked in \cite{neupap} for the boundary sine-Gordon
model.%
}\begin{equation}
R_{\alpha}(\theta)R_{\alpha}(\theta)e^{i2mL\sinh\theta}=1\label{eq:bye1}\end{equation}
 In order to describe the boundary bound states we continue this equation
to imaginary values of the rapidity $\theta$ as in \cite{uv_ir}.
Writing $\theta=iv_{0}+i\delta$ and assuming that $\delta\rightarrow0$
for $L\rightarrow\infty$ one finds \textsl{two} solutions for $\delta$
such that the corresponding $\theta$-s are in the physical strip,
with the following large volume asymptotics:\[
\delta\sim\pm\frac{\Gamma^{2}}{2}e^{-mL\sin v_{0}}\]
 The energy of these two solutions above the ground state is \begin{equation}
E_{\pm}\sim m\cos v_{0}\mp m\frac{\Gamma^{2}}{2}\sin v_{0}\, e^{-mL\sin v_{0}}\label{eq:by_epm}\end{equation}
 The interpretation of this result is clear: the two energies correspond
to the two (approximate) wave functions containing the symmetric (resp.
antisymmetric) combinations of the identical 'half line' bound states
localized in the vicinity of the two boundaries (cf. \cite{uv_ir}).

For solutions of the Bethe-Yang equation (\ref{eq:bye1}) that for
$L\rightarrow\infty$ go to $\theta=i\frac{\pi}{2}$, the asymptotic
behaviour is determined by the residue (\ref{eq:pi2pole}). Furthermore,
writing $\theta=i\frac{\pi}{2}+i\delta$, only the solution with $\delta<0$
is inside the physical strip. Therefore in this case we find the following
result for the energy \[
E_{BY}=m\cosh\theta=m\frac{g_{\alpha}^{2}}{2}e^{-mL}\]
 In contrast to the case of a boundary bound state discussed above,
the other solution is nonphysical. Remembering that $E_{BY}$ is the
energy of the given state relative to the finite volume ground state
of the system, it has a different interpretation than in the previous
case. Since the true ground state is the symmetric combination of
the asymptotic ground states localized at the left/right boundary,
while the excited one is the antisymmetric one \cite{stripcikk},
$E_{BY}$ can be identified with the \textsl{difference} between the
energies of these two states, between which one can switch by changing
the relative sign between the two $\tilde{g}$-s. Using (\ref{eq:Casimir_onepart})
we obtain the relation\[
\Delta E=2m\tilde{g}_{\alpha}^{2}e^{-mL}\equiv E_{BY}=m\frac{g_{\alpha}^{2}}{2}e^{-mL}\]
 Note that formally this result is just half of the energy difference
$E_{+}-E_{-}$ in (\ref{eq:by_epm}) which is due to the special status
of the $i\frac{\pi}{2}$ pole. This is again consistent with eqn.
(\ref{eq:gtilde_g_intro}).

\subsubsection{Connecting the VEV and the boundary state}

Next we consider the connection between the vacuum expectation value
$\,_{\beta}\langle0|\Phi(x,t)|0\rangle_{\beta}$ of a field $\Phi$
which is an interpolating field for the asymptotic particle states
(where $|0\rangle_{\beta}$ denotes the ground state of the half line
$x\leq0$ theory with the boundary condition $\beta$ imposed at $x=0$)
and the coefficient(s) of the expansion of the boundary state, eqn.
(\ref{eq:bstate_onepart}), in the crossed channel. Following the
ideas presented in \cite{dptw_onepoint}, this connection is based
on the fundamental (defining) property of the boundary state, namely
on the equivalence of the correlation functions in the two channels.
Indeed this equivalence applied to the one-point function reads \[
\,\,_{\beta}\langle0|\Phi(x,t)|0\rangle_{\beta}=\langle0|\Phi(y,\tau)|B_{\beta}\rangle\qquad y=it,\quad x=i\tau\]
 Substituting eqn. (\ref{eq:bstate_onepart}) one finds \[
\,\,_{\beta}\langle0|\Phi(x,t)|0\rangle_{\beta}=\langle0|\Phi(0)|0\rangle+\langle0|\Phi(0)|\theta=0\rangle\tilde{g}_{\beta}e^{im\tau}+\dots\]
 where the first term is the v.e.v. of the field in the bulk (which
is an $x$ independent constant) and the dots stand for the contribution
of the multi-particle terms. The matrix element of the field in the
second term is basically the normalization of the bulk one particle
form factor in the closed channel \begin{equation}
\langle0|\Phi(0)|\theta=0\rangle=\sqrt{\frac{Z}{2}}\label{eq:Z_definition}\end{equation}
with $Z$ denoting the wavefunction renormalization constant of $\Phi$.
Exploiting the connection between the coordinates in the two channels
one finds finally the asymptotic ($x\rightarrow-\infty$) expression
\begin{equation}
\,\,_{\beta}\langle0|\Phi(x,t)|0\rangle_{\beta}=\langle0|\Phi(0)|0\rangle+\sqrt{\frac{Z}{2}}\tilde{g}_{\beta}e^{mx}+{\mathcal{O}}(e^{2mx})\label{eq:vevas1}\end{equation}
which gives the following relation \begin{equation}
\bar{g}_{\beta}=\sqrt{\frac{Z}{2}}\tilde{g}_{\beta}\label{eq:gbar_gtilde_relation}\end{equation}

\subsection{(Semi)classical examples}

Most of the examples and considerations presented so far relied on
the integrability of the underlying model. In order to extend the
relations between the various $g$-s to nonintegrable models we now
present some further examples, and also develop methods which do not
need integrability. The general framework in this section is that
of the (semi)classical approximation of the various field theoretical
models.

\subsubsection{Connecting the classical VEV and the ground state energy}

We consider a model described by a scalar field $\Phi$ on a strip
of width $L$ satisfying Dirichlet boundary conditions $\Phi(0,t)=\Phi_{0}$
and $\Phi(L,t)=\Phi_{L}$ with the bulk (Minkowski) action\begin{equation}
\mathcal{A}_{\mathrm{bulk}}=\int dtdx\left\{ \frac{1}{2}\left(\partial_{t}\Phi\right)^{2}-\frac{1}{2}\left(\partial_{x}\Phi\right)^{2}-U\left(\Phi\right)\right\} \label{eq:classical_bulk_action}\end{equation}
The ground state is a solution of the static classical equation of
motion (satisfying also the boundary conditions) \begin{equation}
\frac{1}{2}\left(\frac{\partial\Phi}{\partial x}\right)^{2}-U(\Phi)=C,\qquad U(\Phi)\geq0\label{eq:statice}\end{equation}
 where $U(\Phi)$ describes the self interaction of the scalar field
and $C$ is (an $L$ dependent) constant of integration. For $L\rightarrow\infty$
this constant vanishes ($C\rightarrow0$) and the ground state tends
to a superposition of two 'half line' solutions determined by the
two boundary conditions. This ground state has a simple qualitative
description if we assume that $U(\Phi)$ has a minimum at $\Phi_{*}$
(normalized as $U(\Phi_{*})=0$) and that $\Phi_{0,L}$ are in the
'vicinity' of $\Phi_{*}$ (meaning that no other minimum is between
$\Phi_{*}$ and $\Phi_{0}$ or $\Phi_{L}$): the scalar field starts
at $\Phi_{0}$ then increases (decreases) towards $\Phi_{*}$, reaches
a maximum (minimum) $\Phi_{1}$ close to $\Phi_{*}$, then increases/decreases
to $\Phi_{L}$%
\footnote{This description is valid if $\Phi_{0}$ and $\Phi_{L}$ are both
smaller or both greater than $\Phi_{*}$; if $\Phi_{*}$ is between
them, then $\Phi_{1}=\Phi_{*}$ and the solution is monotonic.%
}. It is important to notice that the $L$ dependence of the ground
state energy can be determined simply from that of the constant $C$.
Indeed, writing \[
E(L)=\int\limits _{0}^{L}dx\left\{ \frac{1}{2}\left(\frac{\partial\Phi}{\partial x}\right)^{2}+U(\Phi)\right\} =-CL+\left\{ \int\limits _{\Phi_{0}}^{\Phi_{1}}+\int\limits _{\Phi_{L}}^{\Phi_{1}}\right\} dv\sqrt{2(U(v)+C)}\]
 and exploiting \[
L=\left\{ \int\limits _{\Phi_{0}}^{\Phi_{1}}+\int\limits _{\Phi_{L}}^{\Phi_{1}}\right\} \frac{dv}{\sqrt{2(U(v)+C)}}\]
 we get \begin{equation}
\frac{dE}{dL}=-C,\qquad\mathrm{thus}\qquad E(L)=E_{\infty}-\int C(L)dL\label{eq:elfugg}\end{equation}
 Therefore we determine next the large $L$ asymptotics of the integration
constant $C(L)$ using the qualitative picture of the ground state.
If, in the half line theory, far away from the boundary, the solutions
with the $\Phi_{0}$ ($\Phi_{L}$) b.c. have the asymptotic form \begin{equation}
\Phi(x)\sim\Phi_{*}+\bar{g}_{0}e^{-mx},\qquad\Phi(x)\sim\Phi_{*}+\bar{g}_{L}e^{-mx}\label{eq:assol1}\end{equation}
 then for large $L$-s, in the \textit{central part} of the strip,
which is far away from both boundaries, the scalar field is given
by (upto subleading exponential corrections) \[
\Phi(x)\sim\Phi_{*}+\bar{g}_{0}e^{-mx}+\bar{g}_{L}e^{m(x-L)}\]
 (Note that the bulk vacuum expectation value $\Phi_{*}$ appearing
in (\ref{eq:assol1}) is a parameter of the bulk theory; the dependence
on the boundary conditions enters only via the coefficients of the
exponential terms). Expanding the scalar potential in the vicinity
of $\Phi_{*}$ as $U(\Phi)=\frac{m^{2}}{2}(\Phi-\Phi_{*})^{2}+{\mathcal{O}}((\Phi-\Phi_{*})^{3})$,
and using eqn. (\ref{eq:statice}) in the \textit{central part} of
the strip, one readily finds \[
C(L)=-2m^{2}\bar{g}_{0}\bar{g}_{L}e^{-mL}\]
 Using this in eqn. (\ref{eq:elfugg}) gives the asymptotic expression
of the ground state energy as \[
E(L)=E_{\infty}-2m\bar{g}_{0}\bar{g}_{L}e^{-mL}\]
 where $E_{\infty}$ is the sum of the energies of the two half line
solutions having the asymptotic behaviour in (\ref{eq:assol1}). This
gives\[
\bar{g}=\frac{\tilde{g}}{\sqrt{2}}\]
which is consistent with eqn. (\ref{eq:gbar_gtilde_relation}) since
for the normalization of the field given in (\ref{eq:classical_bulk_action})
the classical value of the wave function renormalization constant
is $Z=1$.

\subsubsection{Semiclassical limit of sine-Gordon model}

In this subsection we consider the semiclassical limit of sine-Gordon
model on a strip with Dirichlet boundary conditions and determine
explicitely the three characteristic $g$-s in terms of the parameters
of the model, thus verifying the conjectured relations among them.

The bulk Lagrangian is written in eqn. (\ref{eq:sgbulk}) and for
simplicity we consider first the special case when the model is restricted
to the negative half line ($x\leq0$) and the field satisfies the
boundary condition $\Phi(x,t)|_{x=0}=\Phi_{0}^{D}$ where the parameter
$\Phi_{0}^{D}$ is in its fundamental range $0\leq\frac{\beta}{2}\Phi_{0}^{D}\equiv\varphi_{0}\leq\frac{\pi}{2}$.
In this case the (semi)classical ground sate is given by a \lq half'
soliton standing at the location required by the boundary condition
\[
\Phi_{\textrm{bg}}\left(x,a^{+}\right)=\frac{4}{\beta}\arctan\left(e^{m(x-a^{+})}\right)\quad{\textrm{where}}\quad e^{-ma^{+}}=\tan\frac{\varphi_{0}}{2}\]
 Recalling the definition of $\bar{g}$ \[
\Phi_{\textrm{bg}}\sim\bar{g}e^{mx}\quad\mathrm{for}\quad x\rightarrow-\infty\]
 we obtain from the explicit solution\begin{equation}
\bar{g}=\frac{4}{\beta}\tan\frac{\varphi_{0}}{2}\label{eq:gbarsg}\end{equation}

To obtain the semiclassical limit of $g$ coming from the reflection
amplitude one has to determine the appropriate solutions of the differential
equation describing the linearized fluctuations in the standing soliton
background. This was done in \cite{stripcikk}, \cite{KP}, and using
these wave functions one obtains the classical reflection amplitude
\[
R(q)=\frac{m-iq}{m+iq}\,\frac{iq+m\cos\varphi_{0}}{iq-m\cos\varphi_{0}}\,,\qquad q=m\sinh\theta\]
 Surprisingly, $R(q)$ has a second order pole at $\theta=i\frac{\pi}{2}$,
i.e. for $\theta\sim i\frac{\pi}{2}$ it can be written as \[
R(q)\sim-\frac{4\tan^{2}\left(\frac{\varphi_{0}}{2}\right)}{(\theta-i\frac{\pi}{2})^{2}}\]
 We can explain this second order pole in the following way. Since
in the semiclassical limit the elementary field excitations become
identical to the first breather we should compare $R(q)$ to the (semi)classical
limit of the first breather's reflection amplitude (on the ground
state $|\rangle$) (\ref{eq:b1refl}), which has a first order pole
at $\theta=i\frac{\pi}{2}$. However, as shown in \cite{KP}, in the
(semi)classical limit $\lambda\rightarrow\infty$ the $\eta$ boundary
parameter should also be scaled as \[
\eta=\eta_{\mathrm{cl}}(1+\lambda)\]
 keeping $\eta_{\mathrm{cl}}$ finite, which is related to the boundary
value of the field via \[
\eta_{\mathrm{cl}}=\varphi_{0}=\frac{\beta\Phi_{0}^{D}}{2}\]
In addition $R_{|\rangle}^{(1)}(\theta)$ also has a pole at $\hat{\theta}=i\left(\frac{\pi}{2}-\frac{\pi}{2\lambda}\right)$
coming from the $\left(\frac{1}{2\lambda}+\frac{3}{2}\right)$ factor
in the denominator, corresponding to the \textsl{bulk} process of
two $B^{1}$'s fusing into a $B^{2}$ of zero energy (rapidity $\theta=i\pi/2$),
which is then absorbed by the boundary. Clearly for $\beta\rightarrow0$
( $\lambda\rightarrow\infty$) $\hat{\theta}\rightarrow i\frac{\pi}{2}$
and one obtains a second order pole. Therefore for sufficiently small
$\beta$ (sufficiently large $\lambda$) in the vicinity of $\theta\sim i\frac{\pi}{2}$
the quantum reflection amplitude behaves as \[
R_{|\rangle}^{(1)}(\theta)\sim\frac{A}{(\theta-i\frac{\pi}{2})(\theta-i(\frac{\pi}{2}-u_{1}))}\;,\qquad\lim_{\lambda\rightarrow\infty}u_{1}=\lim_{\lambda\rightarrow\infty}\frac{\pi}{2\lambda}=0\]
 In this case the residue of $R_{|\rangle}^{(1)}(\theta)$ at $\theta=i\frac{\pi}{2}$
is \[
\mathop{\mathrm{Res}}_{\theta=i\frac{\pi}{2}}R_{|\rangle}^{(1)}(\theta)=-i\frac{A}{u_{1}}\]
 while in the (semi)classical limit one obtains \[
R(\theta)\sim\frac{A}{(\theta-i\frac{\pi}{2})^{2}}\]
 Combining these expressions with the definition of $g$ one finally
finds \[
i\frac{g^{2}}{2}:=\mathop{\mathrm{Res}}_{\theta=i\frac{\pi}{2}}R_{|\rangle}^{(1)}(\theta)=i\frac{64}{\beta^{2}}\tan^{2}\frac{\varphi_{0}}{2}\]
i.e.\begin{equation}
g=4\sqrt{\frac{8}{\beta^{2}}}\tan\frac{\varphi_{0}}{2}\sim4\sqrt{\frac{\lambda}{\pi}}\tan\frac{\eta}{2\lambda}\label{eq:gsg}\end{equation}
 (where we 'removed' the semiclassical limit in writing the last approximate
equality) which coincides with the semiclassical limit of (\ref{eq:sgg1})
as was found also in \cite{Lusch}.

In \cite{Lusch,MRS} the classical ground state energy of the sine-Gordon
model satisfying the Dirichlet boundary conditions on a strip of width
$L$\[
\Phi(0,t)=\Phi_{0}^{D},\qquad\Phi(L,t)=\Phi_{L}^{D}\]
 (in the sector of zero topological charge) was found to be \[
E(\varphi_{0},\varphi_{L},L)\sim E_{\infty}-\frac{32m}{\beta^{2}}\tan\frac{\varphi_{0}}{2}\tan\frac{\varphi_{L}}{2}e^{-l}\]
in the asymptotic regime $l=mL\gg1$. Here $E_{\infty}$ is the sum
of the energies of the two asymptotic static 'half' solitons representing
the (asymptotic) ground state on the strip. Defining the classical
counterpart of $\tilde{g}$ in (\ref{eq:Casimir_onepart}) as \[
E(L)=E_{\infty}-m\tilde{g}_{0}\,\tilde{g}_{L}e^{-mL}\]
 and comparing to the explicit expression gives \begin{equation}
\tilde{g}_{i}=\frac{4\sqrt{2}}{\beta}\tan\frac{\varphi_{i}}{2}\;,\qquad i=0,\, L\label{eq:ghatsg}\end{equation}
 Combining eqn. (\ref{eq:gbarsg},\ref{eq:gsg}) and (\ref{eq:ghatsg})
one obtains for the semiclassical limit of the various $g$-s: \begin{equation}
\bar{g}_{i}=\frac{g_{i}}{2\sqrt{2}}=\frac{\tilde{g}_{i}}{\sqrt{2}}\;,\qquad i=0,\, L\label{eq:classical_grelation}\end{equation}
 This is consistent with eqn. (\ref{eq:pi2pole}) and (\ref{eq:gtilde_g_intro})
and - since $Z\rightarrow1$ in the (semi)classical limit - also with
eqn. (\ref{eq:vevas1}). 

Similar results were obtained in \cite{fateev_onofri} for the case
of general affine Toda field theories (ATFT) where the authors use
a different normalization for the fields and the particle states than
we do. To facilitate the comparison, we give the relations for the
case $a_{1}^{(1)}$ (sinh-Gordon) here. The action in that paper is
normalized as \[
\mathcal{A}=\int d^{2}x\left\{ \frac{1}{8\pi}\left(\partial_{\mu}\phi\right)^{2}+\mu\left(e^{\sqrt{2}b\phi}+e^{-\sqrt{2}b\phi}\right)\right\} \]
which in our notation means that the classical wave function renormalization
constant is $Z=4\pi$. As a result, from (\ref{eq:Z_definition})
we have\[
\langle0\vert\phi(0)\vert\theta=0\rangle=\sqrt{2\pi}+O(b^{2})\]
while their one-particle state , $\vert\theta\rangle_{\mathrm{FO}}$
is normalized according to \[
\langle0\vert\phi(0)\vert\theta=0\rangle_{\mathrm{FO}}=\sqrt{\pi}+O(b^{2})\]
which means that $\vert\theta=0\rangle=\sqrt{2}\vert\theta=0\rangle_{\mathrm{FO}}$.
The residue of the reflection factor is parameterized in \cite{fateev_onofri}
as \[
R(\theta)\sim\frac{D(b)^{2}}{\theta-i\frac{\pi}{2}}\]
which means that $D(b)$ can be written in our notation as\[
D(b)=\frac{g}{\sqrt{2}}\]
 The form of the one-particle term in the boundary state given in
\cite{fateev_onofri}\[
\vert B\rangle=\left(1+D(b)A_{\mathrm{FO}}^{+}(0)+\dots\right)\vert0\rangle=\left(1+\frac{D(b)}{\sqrt{2}}A^{+}(0)+\dots\right)\vert0\rangle\quad\Rightarrow\quad\tilde{g}=\frac{D(b)}{\sqrt{2}}\]
is then consistent with the relation (\ref{eq:gtilde_g_intro})%
\footnote{We mention, however, that the two-particle term in the cluster expansion
of $\vert B\rangle$ is not properly normalized in \cite{fateev_onofri},
but it plays no role in the issue at hand.%
}. In addition, from the asymptotics of the classical vacuum solution
found in \cite{fateev_onofri} the VEV parameter $\bar{g}$ in the
classical limit $b\rightarrow0$ reads\[
\bar{g}_{\mathrm{classical}}=\frac{1}{b}\lim_{b\rightarrow0}\sqrt{\pi}bD(b)=\sqrt{2\pi}\tilde{g}_{\mathrm{classical}}\]
 which is then consistent with (\ref{eq:gbar_gtilde_intro}). The
results of \cite{fateev_onofri} therefore provide a generalization
of the arguments of the present subsection to the case of any ATFT.

\subsubsection{Semiclassical description of spontaneously broken $\Phi^{4}$ theory}

In this subsection we consider the spontaneously broken $\Phi^{4}$
theory restricted by Dirichlet boundary conditions to a strip of width
$L$. Using semiclassical considerations we express explicitely the
(semiclassical limit of the) three characteristic $g$-s in terms
of the parameters of the model, thus verifying explicitely the conjectured
relations among them in this \textsl{nonintegrable} case. 

The bulk Lagrangian of this model can be written as \begin{equation}
\mathcal{L}=\frac{1}{2}\left(\frac{\partial\Phi}{\partial t}\right)^{2}-\frac{1}{2}\left(\frac{\partial\Phi}{\partial x}\right)^{2}-\frac{\lambda}{4}\left(\Phi^{2}-\frac{m^{2}}{\lambda}\right)^{2}\label{eq:phi4_action}\end{equation}
 where $\lambda>0$ and $m$ is real. Introducing $z=\frac{mx}{\sqrt{2}}$
the well known kink (anti-kink) solution 'standing' at $x=a$ can
be written as \[
\Phi(x)=\pm\frac{m}{\sqrt{\lambda}}\tanh(z-z_{0})\,,\qquad z_{0}=\frac{ma}{\sqrt{2}}\]
 It is also well known that the mass of the elementary particle in
the 'vacuum' sector is $\mu=m\sqrt{2}$.

First consider this model restricted to the half line $x<0$ (or $x>0$)
imposing Dirichlet boundary condition at $x=0$: \[
\Phi(0,t)=\Phi_{0}\]
 which makes it possible to determine the explicit form of two of
the three $g$-s. The essential observation is that -- for $-\frac{m}{\sqrt{\lambda}}<\Phi_{0}<\frac{m}{\sqrt{\lambda}}$
at least -- the (semi)classical ground state of the model, $\Phi_{\textrm{bg}}$,
is given by a bulk kink/anti-kink standing at the position required
by the boundary condition \[
\mp\frac{m}{\sqrt{\lambda}}\tanh(z_{0})=\Phi_{0}\]
 This enables us to determine $\bar{g}$ which is the parameter characterizing
the field in the ground state: choosing, say, the solution that far
away from the boundary at $x=0$ tends to the $\frac{m}{\sqrt{\lambda}}$
bulk ground state, then $\bar{g}$ is defined as \[
\lim_{x\rightarrow-\infty}\Phi_{\textrm{bg}}=\frac{m}{\sqrt{\lambda}}+\bar{g}_{-}e^{\mu x},\qquad\quad\lim_{x\rightarrow\infty}\Phi_{\textrm{bg}}=\frac{m}{\sqrt{\lambda}}+\bar{g}_{+}e^{-\mu x}\]
 depending on whether we restrict the model to the $x<0$ or to the
$x>0$ half line. Using the explicit kink/anti-kink solution one obtains
readily \begin{equation}
\bar{g}_{-}=-\frac{2m}{\sqrt{\lambda}}e^{-2z_{0}},\qquad\bar{g}_{+}=-\frac{2m}{\sqrt{\lambda}}e^{2z_{0}}\label{eq:gbargfi4}\end{equation}

To obtain $g$ which characterizes the singularity of the reflection
amplitude, we restrict the model to the $x<0$ half line and determine
the appropriate solutions of the differential equation describing
the linearized fluctuations in the standing soliton (kink) background.
This differential equation is the same as in the bulk case; the only
remaining task one is to combine the left and right moving bulk wave
solutions to satisfy the boundary condition at $x=0$. In the bulk
case the solutions in the continuum spectrum (which is relevant here)
can be written \cite{DHN} \[
\eta_{q}(z)=e^{iqz}(3\tanh^{2}(z-z_{0})-1-q^{2}-3iq\tanh(z-z_{0}))\]
 where $q$ is a real parameter that determines the frequency of the
fluctuation through $\omega^{2}=m^{2}(2+q^{2}/2)$. We look for solutions
in the form \[
h_{k}(z)=\frac{\eta_{k}(z)}{2-k^{2}-3ik}N_{1}+\frac{\eta_{-k}(z)}{2-k^{2}+3ik}N_{2}\]
 that satisfy the boundary condition $h_{k}(0)=0$ and also for which
the coefficient of $e^{ikz}$ far away from the boundary is one. The
classical limit of the reflection amplitude is then defined by \[
\lim_{z\rightarrow-\infty}h_{k}(z)=e^{ikz}+R_{\textrm{cl}}(k)e^{-ikz}\]
 and from the explicit form of $h_{k}(z)$ one obtains: \[
R_{\textrm{cl}}(k)=-\frac{2-k^{2}-3ik}{2-k^{2}+3ik}\,\frac{3\tanh^{2}(z-z_{0})-1-k^{2}+3ik\tanh(z-z_{0})}{3\tanh^{2}(z-z_{0})-1-k^{2}-3ik\tanh(z-z_{0})}\;,\qquad k=2\sinh\theta\]
 with $\theta$ being the usual rapidity parameter (in terms of which
the spatial momentum of the particle is $p=\mu\sinh\theta$).

\begin{figure}
\begin{center}\includegraphics[%
  scale=0.6]{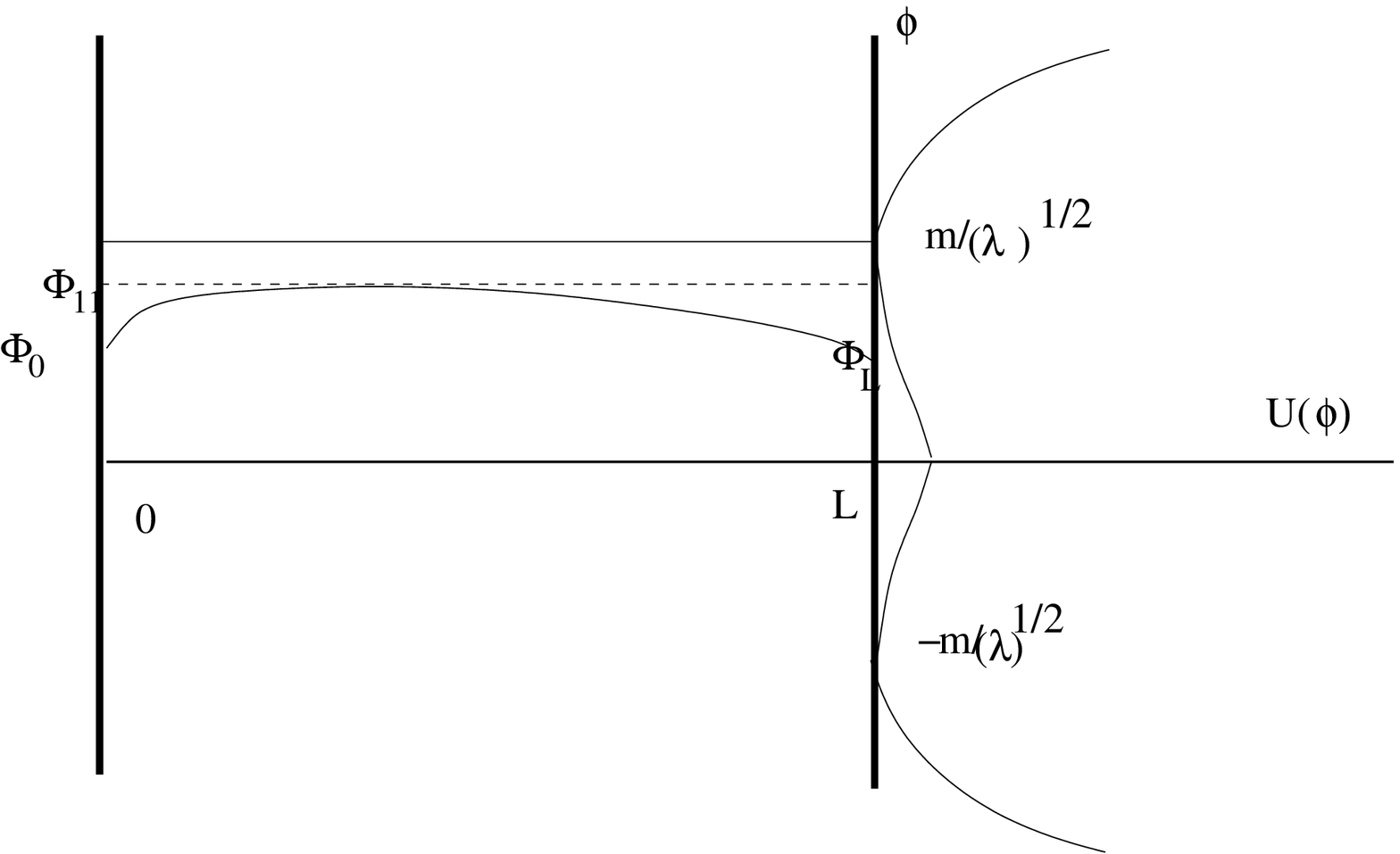}\end{center}

\caption{\label{cap:fig_grstate} Qualitative behaviour of the classical ground
state solution}
\end{figure}

$R_{\textrm{cl}}(k)$ has a second order pole at $\theta=i\frac{\pi}{2}$,
i.e. for $\theta\sim i\frac{\pi}{2}$ it can be written as \[
R_{\textrm{cl}}(k)\sim-\frac{12e^{-4z_{0}}}{(\theta-i\frac{\pi}{2})^{2}}\]
 This second order pole can be explained in the same way as in the
sine-Gordon case. We assume that in the full quantum reflection amplitude
$R_{\textrm{q}}\left(\theta\right)$ there is a pole close to $i\frac{\pi}{2}$
in such a way, that in the semiclassical limit (i.e. when $\lambda\rightarrow0$)
it coincides with the first order 'quantum' pole at $\theta=i\frac{\pi}{2}$:
\[
R_{\textrm{q}}\left(\theta\right)\sim\frac{A}{(\theta-i\frac{\pi}{2})(\theta-i(\frac{\pi}{2}-u))}\;,\qquad\lim_{\lambda\rightarrow0}u=0\]
 Indeed in this case the residue of $R_{\textrm{q}}\left(\theta\right)$
at $\theta=i\frac{\pi}{2}$ is \begin{equation}
\mathop{\mathrm{Res}}_{\theta=i\frac{\pi}{2}}R_{\textrm{q}}\left(\theta\right)=-i\frac{A}{u}\label{eq:g_Au}\end{equation}
 while classically one obtains \[
R_{\textrm{cl}}\left(k\right)\sim\frac{A}{(\theta-i\frac{\pi}{2})^{2}}\]
where \begin{equation}
A=-12e^{-4z_{0}}\label{eq:phi4_A}\end{equation}
This mechanism can only work if in the \textsl{bulk} theory there
is a fusion of appropriate bulk particles to generate the extra pole.
Fortunately in the bulk spontaneously broken $\Phi^{4}$ theory there
is a rather complex semiclassical spectrum of 'approximate breathers'
\cite{DHN} and $u$ can be identified with the fusion angle associated
to the second lightest particle $B_{2}$ as a bound state of two copies
of the lightest particle $B_{1}$: \[
2m_{1}\cos u=m_{2}\]
 Using the semiclassical formulae for $m_{1}$ and $m_{2}$ \cite{DHN}
we get \[
\cos u=\frac{m_{2}}{2m_{1}}\longrightarrow1-\frac{9}{32}\,\frac{\lambda^{2}}{m^{4}}\]
 i.e. \begin{equation}
u=\frac{3}{4}\frac{\lambda}{m^{2}}\label{eq:phi4_u}\end{equation}
 for $\lambda\rightarrow0$. Substituting (\ref{eq:phi4_A},\ref{eq:phi4_u})
into the definition of $g$ gives \begin{equation}
i\frac{g^{2}}{2}=\mathop{\mathrm{Res}}_{\theta=i\frac{\pi}{2}}R_{\textrm{q}}\left(\theta\right)=i\frac{16m^{2}}{\lambda}e^{-4z_{0}},\qquad{\textrm{i.e.}}\qquad g=-m\frac{4\sqrt{2}}{\sqrt{\lambda}}e^{-2z_{0}}\label{eq:gfi4}\end{equation}
 To obtain the explicit form of the third $g$ parameter we confine
the model to a strip of width $L$ by imposing Dirichlet boundary
conditions ($\Phi(0,t)=\Phi_{0}$ and $\Phi(L,t)=\Phi_{L}$) and determine
the classical ground state energy for asymptotically large $L$-s:
\[
E(L)=E_{\infty}-\mu\tilde{g}_{0}\,\tilde{g}_{L}e^{-\mu L}\]
 (In the $\lambda\rightarrow0$ limit the classical energy gives the
leading semiclassical contribution). Assuming $0<\Phi_{0}\,,\Phi_{L}<\frac{m}{\sqrt{\lambda}}$
the qualitative behaviour of the static ground state solution is depicted
in the figure \ref{cap:fig_grstate} (a quantitative description can
be given using the methods of \cite{stripcikk}).

The ground state is a solution of the static classical equation of
motion (\ref{eq:statice}) with $U(\Phi)=\frac{\lambda}{4}(\Phi^{2}-\frac{m^{2}}{\lambda})^{2}$.
To obtain the $L$ dependence of $C$ explicitely we introduce $\Phi(x)=\frac{m}{\sqrt{\lambda}}\varphi(x)$,
$\Phi_{0,1,L}=\frac{m}{\sqrt{\lambda}}\varphi_{0,1,L}$ and $C=-\frac{m^{4}}{4\lambda}d^{2}$,
in terms of which the condition determining $d(L)$ is: \begin{equation}
\frac{m}{\sqrt{2}}L=F(\phi(\varphi_{0})|\alpha)+F(\phi(\varphi_{L})|\alpha)\,,\qquad F(\phi(v)|\alpha)=\int\limits _{v}^{\sqrt{1-d}}\frac{ds}{\sqrt{(1+d-s^{2})(1-d-s^{2})}}\label{ldrel}\end{equation}
 where $F(\phi(v)|\alpha)$ is an elliptic integral of the first kind
with modular angle $\alpha$ and amplitude $\phi(v)$ with \[
\sin\alpha=\sqrt{\frac{1-d}{1+d}}\;,\qquad\sin^{2}\phi=\frac{1+d}{1-d}\,\frac{1-d-v^{2}}{1+d+v^{2}}\]
 For $L\rightarrow\infty$ one has $d\rightarrow0$ and $F$ has the
asymptotic behaviour \[
F(\phi(v)|\alpha)=\frac{1}{2}\ln\frac{8}{d}-\ln\sqrt{\frac{1+v}{1-v}}\]
 Using this in eqn. (\ref{ldrel}) determines the $L$ dependence
of $d$, which finally leads to \[
E(L)=E_{\infty}-8\mu\frac{m^{2}}{\lambda}\frac{1-\varphi_{0}}{1+\varphi_{0}}\,\frac{1-\varphi_{L}}{1+\varphi_{L}}e^{-\mu L}\]
 The boundary conditions require $\tanh z_{0}=-\varphi_{0}$ and $\tanh z_{L}=\varphi_{L}$,
leading to \begin{equation}
\tilde{g}_{0}=-m\frac{2\sqrt{2}}{\sqrt{\lambda}}e^{2z_{0}},\qquad\tilde{g}_{L}=-m\frac{2\sqrt{2}}{\sqrt{\lambda}}e^{-2z_{L}}\label{eq:ghatfi4}\end{equation}
 Combining eqn. (\ref{eq:gbargfi4},\ref{eq:gfi4}) and (\ref{eq:ghatfi4})
one obtains again that the semiclassical limits of the various $g$-s
satisfy (\ref{eq:classical_grelation}), i.e. \[
\bar{g}_{i}=\frac{g_{i}}{2\sqrt{2}}=\frac{\tilde{g}_{i}}{\sqrt{2}}\;,\qquad i=0,L\]
even in this nonintegrable model.

In passing we note that one can consider the $\Phi^{4}$ theory in
the symmetric phase (i.e. without spontaneous symmetry breaking, obtained
formally by changing the sign of $m^{2}$ in (\ref{eq:phi4_action}))
along the same lines as we did in the previous subsections for the
sine-Gordon and for the spontaneously broken $\Phi^{4}$ theories
here. In this case one can determine - using semiclassical considerations
- the $g$ coming from the VEV of the field in the ground state, as
well as the $g$ defined from the ground state energy on the strip
and verify that they satisfy the conjectured relation. For the $g$
determined from the reflection amplitude one encounters the following
problem: although the differential equation of the small fluctuations
around the ground state with the right boundary conditions can be
solved and the classical reflection amplitude obtained, it exhibits
the same second order pole at $\theta=i\frac{\pi}{2}$ as we encountered
in the sine-Gordon and spontaneously broken $\Phi^{4}$ theories.
However, in the symmetric phase of $\Phi^{4}$ theory there is no
analogue of the bulk soliton/kink solutions, and thus neither of their
breather bound states (exact or approximate). Despite this, the 'pole
merging' mechanism described in the previous cases still exists, but
the pole which collides with the one already located at $\theta=i\frac{\pi}{2}$
in the semiclassical limit comes from outside the physical strip,
as appropriate analytic continuation of the reflection factors to
this regime shows.

\section{Boundary state formalism and proof of $g$-$\bar{g}$-$\tilde{g}$
relations in $D+1$ dimensions}

In this section we analyze quantum field theories in $D+1$ dimensions
in the presence of a $D$ dimensional flat boundary. The correlators
are defined via the path integral approach, and are then expressed
using two alternative Hamiltonian descriptions of the system. If the
boundary is in space ('open' channel) the information on the boundary
condition is encoded in the reflection factors, which corresponds
to the on-shell part of the correlators. However, when the boundary
is situated in time ('closed' channel), the information is contained
in the boundary state. By deriving a reduction formula in the 'closed'
channel we are able to relate the matrix elements of this state to
the on-shell part of the correlators and thus to the reflection factors.
The knowledge of the boundary state, especially the one-particle boundary
coupling, makes it possible to calculate the large distance behaviour
of the one-point function, as well as to determine the leading finite
size corrections of the ground state energy (Casimir effect). Using
the clustering property of the two-point function, these two quantities
can be related to the singular part of the one-particle reflection
amplitude. In this section we focus our attention on one- and two-point
functions; the analysis of the multi-pont correlation functions is
relegated to the appendix.

\subsection{The concept of the boundary state}

Consider an Euclidean quantum field theory of a scalar field $\Phi$
defined in a $D+1$ dimensional half spacetime, parameterized as $(x\leq0,y,\vec{r})$,
in the presence of a codimension one flat boundary located at $x=0.$
The correlation functions defined as \begin{equation}
\langle\Phi(x_{1},y_{1},\vec{r}_{1})\dots\Phi(x_{N},y_{N},\vec{r}_{N})\rangle=\frac{\int\,\mathcal{D}\Phi\:\Phi(x_{1},y_{1},\vec{r}_{1})\dots\Phi(x_{N},y_{N},\vec{r}_{N})\, e^{-S[\Phi]}}{\int\,\mathcal{D}\Phi\, e^{-S[\Phi]}}\label{correlator}\end{equation}
contain all information about the theory. The measure in the functional
integral is provided by the classical action\[
S[\Phi]=\int d\vec{r}\int_{-\infty}^{\infty}dy\left[\int_{-\infty}^{0}dx\left(\frac{1}{2}(\partial_{x}\Phi)^{2}+\frac{1}{2}(\partial_{y}\Phi)^{2}+\frac{1}{2}(\vec{\partial}\Phi)^{2}+U(\Phi)\right)+U_{B}(\Phi(x=0,y,\vec{r}))\right]\]
 which determines also the boundary condition via the boundary potential
$U_{B}$. 

This Euclidean quantum field theory can be considered as the imaginary
time version of two different Minkowskian quantum field theories.
We can consider $y=it$ as the imaginary time and so the boundary
is located in space providing nontrivial boundary condition for the
field $\Phi.$ We refer to this description as the \emph{open channel}.
The field $\Phi$ is an operator valued distribution that satisfies
the equal time commutation relation\[
[\Phi(x,t,\vec{r}),\partial_{t}\Phi(x^{'},t,\vec{r}\:')]=i\delta(x-x^{'})\delta(\vec{r}-\vec{r}\:')\quad;\qquad x,x^{'}<0\]
The space of states in this Hamiltonian description is the boundary
Hilbert space $\mathcal{H}_{B}$ determined by the configurations
on the equal time slices. $\mathcal{H}_{B}$ contains multi-particle
states and is built over the boundary vacuum $\vert0\rangle_{B}$
by the successive application of the particle creation operators%
\footnote{One can also introduce particle-like excitations confined to the boundary
\cite{BBT}, but here we do not consider them for simplicity.%
}. In the asymptotic past the particles do not interact and behave
as free particles travelling towards the boundary; thus \[
\mathcal{H}_{B}=\left\{ a_{in}^{+}(k_{1},\vec{k}_{1})\dots a_{in}^{+}(k_{N},\vec{k}_{N})\vert0\rangle_{B}\quad,\qquad k_{1}\geq\dots\geq k_{N}>0\right\} \]
where the operator $a_{in}^{+}(k,\vec{k})$, normalized as \[
[a_{in}(k,\vec{k}),\, a_{in}^{+}(k^{'},\vec{k}\:')]=(2\pi)^{D}\omega(k,\vec{k})\delta(k-k^{'})\delta(\vec{k}-\vec{k}\:')\quad;\qquad k,k^{'}>0\]
creates an asymptotic particle of mass $m$ with transverse momentum
$k$ and parallel momentum $\vec{k}$. The corresponding energy is
$\omega(k,\vec{k})=\sqrt{k^{2}+\vec{k}^{2}+m^{2}}=\sqrt{k^{2}+m_{\mathrm{eff}}(\vec{k})^{2}}$.
In the Heisenberg picture the time evolution of the field\[
\Phi(x,t,\vec{r})=e^{iH_{B}t}\Phi(x,0,\vec{r})e^{-iH_{B}t}\]
 is generated by the following boundary Hamiltonian \[
H_{B}=\int d\vec{r}\left[\int_{-\infty}^{0}dx\left(\frac{1}{2}\Pi_{t}^{2}+\frac{1}{2}(\partial_{x}\Phi)^{2}+\frac{1}{2}(\vec{\partial}\Phi)^{2}+U(\Phi)\right)+U_{B}(\Phi(x=0))\right]\]
The correlator (\ref{correlator}) can then be understood as the matrix
element \[
\langle\Phi(x_{1},y_{1},\vec{r}_{1})\dots\Phi(x_{N},y_{N},\vec{r}_{N})\rangle=\,_{B}\langle0\vert T_{t}\left(\Phi(x_{1},t_{1},\vec{r}_{1})\dots\Phi(x_{N},t_{N},\vec{r}_{N})\right)\vert0\rangle_{B}\]
where $T_{t}$ denotes time ordering with respect to time $t$, and
the vacuum $\vert0\rangle_{B}$ is normalized to $1$.

Alternatively we can consider $x=i\tau$ as the Minkowskian time.
In this case the boundary is located in time and we can use the usual
infinite volume Hamiltonian description. This is referred to as the
\emph{closed channel}. The Hilbert space is the bulk Hilbert space
$\mathcal{H}$ spanned by multi-particle \emph{in} states \[
\mathcal{H}=\left\{ A_{in}^{+}(\kappa_{1},\vec{k}_{1})\dots A_{in}^{+}(\kappa_{N},\vec{k}_{N})\vert0\rangle\quad,\qquad k_{1}\geq\dots\geq k_{N}\right\} \]
 where the particle creation operators are normalized as \[
[A_{in}(\kappa,\vec{k}),\, A_{in}^{+}(\kappa^{'},\vec{k}\:')]=(2\pi)^{D}\omega(\kappa,\vec{k})\delta(\kappa-\kappa^{'})\delta(\vec{k}-\vec{k}\:')\]
Time evolution \[
\Phi(\tau,y,\vec{r})=e^{iH\tau}\Phi(0,y,\vec{r})e^{-iH\tau}\]
is generated by the bulk Hamiltonian \[
H=\int d\vec{r}\int_{-\infty}^{\infty}dy\left(\frac{1}{2}\Pi_{\tau}^{2}+\frac{1}{2}(\partial_{y}\Phi)^{2}+\frac{1}{2}(\vec{\partial}\Phi)^{2}+U(\Phi)\right)\]
 The boundary appears in time as a final state in calculating the
correlator (\ref{correlator}): \[
\langle\Phi(x_{1},y_{1},\vec{r}_{1})\dots\Phi(x_{N},y_{N},\vec{r}_{N})\rangle=\langle B\vert T_{\tau}\left(\Phi(\tau_{1},y_{1},\vec{r}_{1})\dots\Phi(\tau_{N},y_{N},\vec{r}_{N})\right)\vert0\rangle\]
The state $\langle B\vert$ is called the boundary state, which is
an element of the bulk Hilbert space and is defined by the equality
of the two alternative Hamiltonian descriptions \[
\langle B\vert T_{\tau}\left(\Phi(\tau_{1},y_{1},\vec{r}_{1})\dots\Phi(\tau_{N},y_{N},\vec{r}_{N})\right)\vert0\rangle=\,_{B}\langle0\vert T_{t}\left(\Phi(x_{1},t_{1},\vec{r}_{1})\dots\Phi(x_{N},t_{N},\vec{r}_{N})\right)\vert0\rangle_{B}\]
 where the correspondence is valid if $(i\tau,y)$ is identified with
$(x,it)$. Using asymptotic completeness the boundary state can be
expanded in the basis of asymptotic \emph{in} states as \begin{equation}
\langle B\vert=\langle0\vert\left\{ 1+\bar{K}^{1}A_{in}(0,0)+\int_{0}^{\infty}\frac{d\kappa}{2\pi}\int\frac{d\vec{k}}{(2\pi)^{D-1}\omega(\kappa,\vec{k})}\bar{K}^{2}(\kappa,\vec{k})A_{in}(-\kappa,-\vec{k})A_{in}(\kappa,\vec{k})+\dots\right\} \label{eq:bstate_expanded}\end{equation}
which we refer to as the cluster expansion for the boundary state
(where due to translational invariance only bulk multi-particle states
with zero total momentum can appear). The bars on top of the $K$
coefficients indicate that the above expansion is that of the conjugate
boundary state.

\subsection{One-point function and $K^{1}$}

The one-point function of the field, due to unbroken Poincar\'e symmetry
along the boundary, only has a nontrivial dependence on $x$ and can
be written (using the 'open' channel formulation) as\[
\,_{B}\langle0\vert\Phi(x,t,\vec{r})\vert0\rangle_{B}=G_{bdry}^{1}(x)\]
 Going over to momentum space by Fourier transformation \[
\,_{B}\langle0\vert\Phi(x,t,\vec{r})\vert0\rangle_{B}=\int\frac{d\omega}{2\pi}\int\frac{d\vec{k}}{(2\pi)^{D-1}}\int\frac{dk}{2\pi}e^{i(\omega t-kx-\vec{k}\vec{r})}G_{bdry}^{1}(\omega,k,\vec{k})\]
and analyzing the all order perturbative expression the following
form is obtained:\[
G_{bdry}^{1}(\omega,k,\vec{k})=(2\pi)^{D}\delta(\omega)\delta(\vec{k})G_{bulk}^{2}(\omega,k,\vec{k})\left[2\pi\delta(k)G_{bulk}^{1}+B_{bdry}^{1}(k)\right]\]
Unbroken translational invariance along the boundary is manifested
in the conservation of energy and parallel momentum. Summing up the
contributions of the outer leg the bulk two-point function, a pre-factor
$G_{bulk}^{2}$ can always be isolated. The remaining factor has a
part that preserves the transverse momentum (which is the same as
the bulk one-point function $G_{bulk}^{1}$) and another one which
depends on the boundary condition and violates transverse momentum
conservation. The bulk two-point function has the well-known K\"allen-Lehmann
representation in terms of the spectral function $\sigma(m)$ \[
G_{bulk}^{2}(\omega,k,\vec{k})=\frac{iZ}{\omega^{2}-k^{2}-\vec{k}^{2}-m^{2}+i\epsilon}+\int_{2m}^{\infty}dm^{'}\frac{i\sigma(m^{'})}{\omega^{2}-k^{2}-\vec{k}^{2}-m^{'2}+i\epsilon}\]
where the wave function renormalization constant $Z$ characterizes
the strength of the on-shell part. Performing the energy-momentum
integration and picking up the pole terms the one-point function can
be expressed as \begin{equation}
\,_{B}\langle0\vert\Phi(x,t,\vec{r})\vert0\rangle_{B}=\langle0\vert\Phi(0)\vert0\rangle-\frac{iZ}{2m}B_{bdry}^{1}(im)e^{mx}-\int_{2m}^{\infty}dm^{'}\frac{i\sigma(m^{'})}{2m^{'}}B_{bdry}^{1}(im^{'})e^{m^{'}x}\label{VEV}\end{equation}
From this we can read off the leading large distance behaviour (dominated
by the on-shell part):\begin{equation}
\,_{B}\langle0\vert\Phi(x,t,\vec{r})\vert0\rangle_{B}=\langle0\vert\Phi(0)\vert0\rangle+\bar{g}e^{mx}\quad;\qquad\bar{g}=-\frac{iZ}{2m}B_{bdry}^{1}(im)\label{eq:gbar_B1}\end{equation}
Our aim now is to connect the quantity $\bar{g}$ to the coefficient
$K^{1}$ of the boundary state (\ref{eq:bstate_expanded}). This can
be accomplished by calculating the matrix element\[
\langle B\vert A_{in}^{+}(k,\vec{k})\rangle=(2\pi)^{D}\omega(\kappa,\vec{k})\delta(\kappa)\delta(\vec{k})\bar{K}^{1}\]
in the closed channel. Using the reduction formula derived in the
appendix, it can be expressed in terms of the one-point function as
\begin{eqnarray*}
\langle B\vert A_{in}^{+}(\kappa,\vec{k})\rangle & = & \frac{i}{\sqrt{2Z}}\int_{-\infty}^{\infty}dy\int_{-\infty}^{0}d\tau\int d\vec{r}\, e^{-i\omega(\kappa,\vec{k})\tau+i\kappa y+i\vec{k}\vec{r}}\\
 &  & \hspace{1cm}\left\{ \partial_{\tau}^{2}-\partial_{y}^{2}-\vec{\partial}^{2}+m^{2}-\delta(\tau)(\partial_{\tau}+i\omega(\kappa,\vec{k}))\right\} \,\langle B\vert\Phi(\tau,y,\vec{r})\vert0\rangle\end{eqnarray*}
In the appendix it is also shown that only the on-shell part of the
correlator contributes to the quantities $K$, and thus one can substitute
\[
\langle B\vert\Phi(\tau,y,\vec{r})\vert0\rangle=\,_{B}\langle0\vert\Phi(x=i\tau,t=-iy,\vec{r})\vert0\rangle_{B}\approx-\frac{iZ}{2m}e^{im\tau}B_{bdry}^{1}(im)\]
Plugging this expression back we obtain \[
\bar{K}^{1}=\frac{-i}{m}\sqrt{\frac{Z}{2}}\left[G_{boundary}^{1}(im)\right]=\tilde{g}\]
from which using (\ref{eq:gbar_B1}) we can establish the relation
\[
\bar{g}=\sqrt{\frac{Z}{2}}\tilde{g}\]
We note that the derivation remains valid if the Lagrangian field
$\Phi$ is replaced by any bulk interpolating field for the asymptotic
particles and its appropriate wave function renormalization $Z$.

\subsection{Two-point function and the relation between $R_{1}^{1}$ and $K^{2}$}

The two-point function in the open channel (using the unbroken spacetime
symmetries) can be written as \begin{eqnarray*}
\,_{B}\langle0\vert T(\Phi(x,t,\vec{r})\Phi(x^{'},t^{'},\vec{r}\:^{'}))\vert0\rangle_{B} & =\\
 &  & \hspace{-3.5cm}\int\frac{d\omega}{2\pi}\int\frac{d\vec{k}}{(2\pi)^{D-1}}\int\frac{dk}{2\pi}\int\frac{dk^{'}}{2\pi}e^{i\omega(t-t^{'})-ikx-ik^{'}x^{'}-i\vec{k}(\vec{r}-\vec{r}\,^{'})}\tilde{G}_{bdry}^{2}(\omega,k,k^{'},\vec{k})\end{eqnarray*}
after integrating out $\omega'$ and $\vec{k}'$ using the delta-functions
in (\ref{Greenf}), where \[
\tilde{G}_{bdry}^{2}(\omega,k,k^{'},\vec{k})=2\pi\delta(k-k^{'})G_{bulk}^{2}(\omega,k,\vec{k})+G_{bulk}^{2}(\omega,k,\vec{k})B_{bdry}^{2}(\omega,k,k^{'},\vec{k})G_{bulk}^{2}(\omega,k^{'},\vec{k})\]
 Its on-shell part determines the two-particle reflection factor,
which is the probability amplitude of a transition from a one-particle
initial state $\vert k^{'},\vec{k}^{'}\rangle_{in}=a_{in}^{+}(k,\vec{k})\vert0\rangle_{B}$
into a one-particle final state $\vert k,\vec{k}\rangle_{out}=a_{out}^{+}(k,\vec{k})\vert0\rangle_{B}$
and can be expressed as the matrix element: \[
_{out}\langle k,\vec{k}\vert k^{'},\vec{k}\:^{'}\rangle_{in}=(2\pi)^{D}\omega(k,\vec{k})\delta(k-k^{'})\delta(\vec{k}-\vec{k}\:^{'})R_{1}^{1}(\omega(k,\vec{k}),k)\]
where $R_{1}^{1}$ is the one-particle to one-particle reflection
factor (cf. (\ref{Rkl})). In order to simplify notation (and to conform
better with the conventions of section 2) we drop the indices and
denote $R_{1}^{1}$ simply by $R$ from now on. The boundary reduction
formula connects it to the correlator as%
\footnote{The normalization of the creation operators agrees with \cite{Casimir}
but differs by a factor $\sqrt{2}$ compared to \cite{BBT} and this
affects the reduction formulae.%
} \begin{eqnarray*}
_{out}\langle k,\vec{k}\vert k^{'},\vec{k}\:^{'}\rangle_{in} & = & _{out}\langle k,\vec{k}\vert k^{'},\vec{k}\:^{'}\rangle_{out}\\
 &  & \hspace{-3cm}-2Z^{-1}\int_{-\infty}^{0}dx\int dt\int d\vec{r}\, e^{-i\omega(k,\vec{k})t+i\vec{k}\vec{r}}\cos(kx)\left\{ \partial_{t}^{2}-\partial_{x}^{2}-\vec{\partial}^{2}+m^{2}+\delta(x)\partial_{x}\right\} \\
 &  & \hspace{-2cm}\int_{-\infty}^{0}dx\int dt\int d\vec{r}\, e^{-i\omega(k^{'},\vec{k}\:^{'})t^{'}+i\vec{k}\:^{'}\vec{r}^{'}}\cos(k^{'}x^{'})\left\{ \partial_{t^{'}}^{2}-\partial_{x^{'}}^{2}-\vec{\partial}^{'2}+m^{2}+\delta(x^{'})\partial_{x^{'}}\right\} \\
 &  & \hspace{6cm}\,_{B}\langle0\vert T_{t}(\Phi(x,t,\vec{r})\Phi(x^{'},t^{'},\vec{r}\:^{'}))\vert0\rangle_{B}\end{eqnarray*}
As was explained in \cite{BBT} and is shown in the appendix only
the on-shell part contributes. As a consequence we can keep the on-shell
part from the bulk two-point function and represent the correlator
equivalently as \begin{eqnarray*}
\,_{B}\langle0\vert T_{t}(\Phi(x,t,\vec{r})\Phi(x^{'},t^{'},\vec{r}\:^{'}))\vert0\rangle_{B} & \approx\\
 &  & \hspace{-6cm}\int\frac{d\omega}{2\pi}\frac{dk}{2\pi}\frac{d\vec{k}}{(2\pi)^{D-1}}\frac{iZe^{-i\omega(t-t^{'})}e^{i\vec{k}(\vec{r}-\vec{r}\:^{'})}}{\omega^{2}-k^{2}-\vec{k}^{2}-m^{2}+i\epsilon}\left[e^{-ik(x-x^{'})}+\frac{Z}{2k}B_{bdry}^{2}(\omega,k,k,\vec{k})e^{-ik(x+x^{'})}\right]\end{eqnarray*}
 Plugging this expression into the reduction formula, the bulk part
of the two-point function cancels the disconnected piece and the reflection
factor turns out to be \[
R(\omega(k,\vec{k}),k)=\frac{Z}{2k}B_{bdry}^{2}(\omega(k,\vec{k}),k,k,\vec{k})\]
Let us connect this quantity to the quantity $K^{2}$ in the boundary
state (\ref{eq:bstate_expanded}), considering the following matrix
element: \[
\langle B\vert A_{in}^{+}(\kappa,\vec{k})A_{in}^{+}(\kappa^{'},\vec{k}\:^{'})\rangle=(2\pi)^{D}\omega(\kappa,\vec{k})\delta(\kappa+\kappa^{'})\delta(\vec{k}+\vec{k}\:^{'})\bar{K}^{2}(\kappa,\vec{k})\]
where $\kappa>\kappa^{'}$ is assumed. Applying the result of the
reduction formula presented in the appendix we have\begin{eqnarray*}
\langle B\vert A_{in}^{+}(\kappa,\vec{k})a_{in}^{+}(A^{'},\vec{k}\:^{'})\rangle & =\\
 &  & \hspace{-5.5cm}-(2Z)^{-1}\int_{-\infty}^{\infty}dy\int_{-\infty}^{0}d\tau\int d\vec{r}\, e^{-i\omega(\kappa,\vec{k})\tau+i\kappa y+i\vec{k}\vec{r}}\left\{ \partial_{\tau}^{2}-\partial_{y}^{2}-\vec{\partial}\,^{2}+m^{2}-\delta(\tau)(i\omega(\kappa,\vec{k})+\partial_{\tau})\right\} \\
 &  & \hspace{-5.5cm}\int_{-\infty}^{\infty}dy^{'}\int_{-\infty}^{0}d\tau^{'}\int d\vec{r}\:'\, e^{-i\omega(\kappa^{'},\vec{k}\:^{'})\tau^{'}+ik^{'}y^{'}+i\vec{k}\:^{'}\vec{r}\:^{'}}\left\{ \partial_{\tau^{'}}^{2}-\partial_{x^{'}}^{2}-\vec{\partial^{'}}^{2}+m^{2}-\delta(\tau^{'})(i\omega(\kappa^{'},\vec{k}\:^{'})+\partial_{\tau^{'}})\right\} \,\\
 &  & \hspace{5cm}\langle B\vert T_{\tau}(\Phi(\tau,y,\vec{r})\Phi(\tau^{'},y^{'},\vec{r}\:^{'})\vert0\rangle\end{eqnarray*}
Using the on-shell part of the two-point function and exchanging the
role of space and time as $x=i\tau$ and $y=it$ we obtain \begin{eqnarray*}
\langle B\vert T_{\tau}(\Phi(\tau,y,\vec{r})\Phi(\tau^{'},y^{'},\vec{r}\:')\vert0\rangle & = & \,_{B}\langle0\vert T_{t}(\Phi(x,t,\vec{r})\Phi(x^{'},t^{'},\vec{r}\:'))\vert0\rangle_{B}\approx\\
 &  & \hspace{-6cm}\int\frac{d\omega}{2\pi}\frac{dk}{2\pi}\frac{d\vec{k}}{(2\pi)^{D-1}}\frac{iZe^{ik(y-y^{'})}e^{i\vec{k}(\vec{r}-\vec{r}\:')}}{\omega^{2}-k^{2}-\vec{k}^{2}-m^{2}+i\epsilon}\left[e^{i\omega(\tau-\tau^{'})}+\frac{Z}{2i\omega}B_{bdry}^{2}(-ik,i\omega,i\omega,\vec{k})e^{i\omega(\tau+\tau^{'})}\right]\end{eqnarray*}
where additionally we also exchanged the role of the energy and transverse
momentum as $\omega\leftrightarrow-ik$. In the reduction formula
the bulk part depending on $\tau-\tau^{'}$ does not contribute and
we obtain\[
\bar{K}^{2}(\kappa,\vec{k})=\frac{Z}{2i\omega(\kappa,\vec{k})}B_{bdry}^{2}(-i\kappa,i\omega(\kappa,\vec{k}),i\omega(\kappa,\vec{k}),\vec{k})\]
which means that the relation between $\bar{K}^{2}$ and $R$ is \[
\bar{K}^{2}(\kappa,\vec{k})=R(\omega\to-i\kappa,k\to i\omega)\]
In two spacetime dimensions one can use the rapidity parametrization\[
\omega=m\cosh\vartheta\quad,\quad\kappa=m\sinh\vartheta\]
and then\[
\bar{K}^{2}\left(\vartheta\right)=R\left(i\frac{\pi}{2}+\vartheta\right)\]
which is the same as the relation derived by Ghoshal and Zamolodchikov
\cite{gz}.

\subsection{The connection between $\bar{g}$ and the reflection factor }

The singularity property of the two-particle reflection factor generally
follows from unitarity, similarly to the well-known bulk theory of
the analytic $S$ matrix. Unitarity of the reflection operator $\mathcal{R}$
can be expressed in terms of the interaction matrix $\mathcal{T}$
as \[
\mathcal{R}\mathcal{R}^{+}=(1+i\mathcal{T})(1-i\mathcal{T}^{+})=1\]
Calculating the one-particle matrix element of this relation between
states $\vert k,\vec{k}\rangle_{B}$ and $\vert k^{'},\vec{k}^{'}\rangle_{B}$
and inserting a resolution of the identity we have \begin{eqnarray*}
2\Im m\mathcal{T}(k,\vec{k})(2\pi)^{D}\omega(k,\vec{k})\delta(k-k^{'})\delta(\vec{k}-\vec{k}^{'}) & = & {}_{B}\langle k,\vec{k}\vert\mathcal{T}\vert0\rangle_{B}\,_{B}\langle0\vert\mathcal{T}^{+}\vert k^{'},\vec{k}^{'}\rangle_{B}+\\
 &  & \hspace{-6cm}\sum_{n}\int\frac{d\vec{q}}{(2\pi)^{D-1}\omega(iu_{n}\left(\vec{q}\right),\vec{q})}{}_{B}\langle k,\vec{k}\vert\mathcal{T}\vert iu_{n}\left(\vec{q}\right),\vec{q}\rangle_{B}\,_{B}\langle iu_{n}\left(\vec{q}\right),\vec{q}\vert\mathcal{T}^{+}\vert k^{'},\vec{k}^{'}\rangle_{B}+\\
 &  & \hspace{-6cm}\int\frac{d\vec{q}}{(2\pi)^{D-1}}\int_{0}^{\infty}\frac{dq}{2\pi\omega(q,\vec{q})}{}_{B}\langle k,\vec{k}\vert\mathcal{T}\vert q,\vec{q}\rangle_{B}\,_{B}\langle q,\vec{q}\vert\mathcal{T}^{+}\vert k^{'},\vec{k}^{'}\rangle_{B}+\dots\end{eqnarray*}
where the second term on the right hand side corresponds to boundary
bound states (degrees of freedom propagating along the boundary) which
are pole singularities located at $k=iu_{n}\left(\vec{k}\right)$
with $0<u_{n}(\vec{k})<m$. Unitarity (as expressed above) gives the
following relation for the residue of the reflection factor: \begin{eqnarray*}
2\Im m\left[\omega(iu,\vec{k})R(iu,0)\right] & = & 2\pi\frac{\omega(iu,0)}{iu}\delta(k-iu)C_{n}\left(\vec{k}\right)C_{n}\left(\vec{k}\right)^{\dagger}\end{eqnarray*}
where $C_{n}\left(\vec{k}\right)$ is the strength of the on-shell
vertex corresponding to the creation of the boundary bound state by
the bulk particle. 

Using the unbroken spacetime symmetries we can extract energy and
parallel momentum conservation which in the first term gives \[
{}_{B}\langle k,\vec{k}\vert\mathcal{T}\vert0\rangle_{B}={}_{B}\langle k,\vec{k}\vert\mathcal{R}\vert0\rangle_{B}=(2\pi)^{D}\delta(\vec{k})\delta(\omega(k,\vec{k}))R^{1}(im)\]
This term vanishes at physical energies but we can try to make an
analytical continuation via the boundary reduction formula and express
$R^{1}$ from the one-point function. The energy delta function can
be rewritten as \[
\delta(\omega(k,\vec{0}))=\frac{\omega(k,\vec{0})}{k}\delta(k-im)\]
 and the kinematical pre-factor vanishes exactly at the momentum value
it is concentrated on. This shows that the singularity of the two-particle
reflection matrix at $k=im$ and $\vec{k}=0$ is not a pole in the
momentum variable $k$. We see that in this case the unitarity argument
does not connect the singular part of the reflection factor to the
one-particle emission amplitude and so to the on-shell part of the
one-point function. The fact that there is an essential difference
between the behaviour of this particular type of singularity and the
ones associated to boundary bound states is consistent with their
different treatment in the Bethe-Yang analysis presented in subsection
2.1.3. Therefore we choose an alternative way to connect the strength
of the singularity in the reflection factor to the VEV of the field,
making use of the clustering property of the two-point function. 

In the Euclidean regime the on-shell part of the two-point function,
which determines its behaviour in the asymptotic regime far from the
boundary, reads as 

\begin{eqnarray}
\,_{B}\left\langle 0\right|\Phi(x,y,\vec{r})\Phi(x^{'},y^{'},\vec{r}^{'})\left|0\right\rangle _{B} & \approx\label{eq:euclideanpropagator}\\
 &  & \hspace{-5cm}\int\frac{d\rho}{2\pi}\frac{dk}{2\pi}\frac{d\vec{k}}{(2\pi)^{D-1}}\frac{Z}{\rho^{2}+k^{2}+\vec{k}^{2}+m^{2}}\mathrm{e}^{-i\rho(y-y^{'})}e^{i\vec{k}(\vec{r}-\vec{r}^{'})}\left\{ \mathrm{e}^{-ik(x-x^{'})}+R(k,\vec{k})\mathrm{e}^{-ik(x+x^{'})}\right\} \nonumber \end{eqnarray}
where we supposed $t>t^{'}$ and introduced $\omega=-i\rho$, $t=iy$.
Cluster property implies that for large temporal separation the Euclidean
two-point function satisfies\[
\,_{B}\left\langle 0\right|\Phi(x,y,\vec{r})\Phi(x^{'},y^{'},\vec{r}^{'})\left|0\right\rangle _{B}=\,_{B}\left\langle 0\right|\Phi(x,y,\vec{r})\left|0\right\rangle _{B}\,\,_{B}\left\langle 0\right|\Phi(x^{'},y^{'},\vec{r}^{'})\left|0\right\rangle _{B}+\textrm{O}\left(\mathrm{e}^{-\mu\vert y-y^{'}\vert}\right)\]
with some characteristic scale $\mu$ that corresponds to the gap
in the spectrum above the vacuum. The presence of the disconnected
piece signals the nontrivial vacuum expectation value of the field
$\Phi$ (which is time independent $\partial_{y}\left\langle 0\right|\Phi(x,y,\vec{r})\left|0\right\rangle =0$
due to $y$-translational invariance, and is also independent of $\vec{r}$
for a similar reason). In the asymptotic regime we then expect the
following behaviour\begin{equation}
\,_{B}\left\langle 0\right|\Phi(x,y,\vec{r})\Phi(x^{'},y^{'},\vec{r}^{'})\left|0\right\rangle _{B}\sim\tilde{g}^{2}\mathrm{e}^{m(x+x^{'})}\qquad\vert y-y^{'}\vert\rightarrow\infty\label{eq:gtildecluster}\end{equation}
In a free theory, with $Z=1$ and $R(k,\vec{k})=\pm1$ (corresponding
to Neumann/Dirichlet boundary conditions), the integral (\ref{eq:euclideanpropagator})
can be evaluated explicitely with the result \[
\frac{1}{2\pi}\left(K_{0}\left(mr_{-}\right)\pm K_{0}\left(mr_{+}\right)\right)\;,\quad r_{\pm}=\sqrt{(y-y')^{2}+(x\pm x')^{2}+(\vec{r}-\vec{r}^{'})^{2}}\]
which decays exponentially when $\vert y-y^{'}\vert\rightarrow\infty$
and so there is no disconnected piece. In fact, this remains true
as long as $R(k,\vec{k})$ is regular at $k=\pm im$ as will be implied
by the analysis below.

Here we have to make an Ansatz for the singularity type of the reflection
factor. The expectation from its structure is that it provides the
needed clustering, moreover it should match with the exact results
available in two dimensions. In two dimensional integrable theories
the singularity of the reflection factor at $\vartheta=\frac{i\pi}{2}$
is a pole in the rapidity variable ($k=m\sinh\theta)$ of the form
\[
R(\vartheta)\sim\frac{ig^{2}/2}{\vartheta-i\frac{\pi}{2}}\sim-\frac{g^{2}/2}{\cosh\vartheta}\]
Changing the rapidity variable to the momentum this corresponds to
the behaviour\[
R(k)\sim-\frac{mg^{2}/2}{\sqrt{k^{2}+m^{2}}}\]
Observe that the singularity in the variable $k$ is not of a pole
type, but is milder, just as expected from the unitarity argument.
In higher dimensions we expect this singularity to be connected with
the virtual one-particle emission (just as in the 1+1 dimensional
case), and due to parallel momentum conservation it is expected to
have the following form:\begin{equation}
R(k,\vec{k})\sim-\frac{mg^{2}/2}{\sqrt{k^{2}+\vec{k}^{2}+m^{2}}}(2\pi)^{D}\delta(\vec{k})\label{eq:2pt_sing_assumption}\end{equation}
 Such a term introduces a singularity at $k^{2}=-m^{2}$ where the
on-shell value of $\rho$ is $0$ which means a contribution which
is constant in the temporal separation $\vert y-y^{'}\vert$. We remark
that in order to obtain this behaviour from the analytic R-matrix
theory the classification and analysis of Coleman-Thun diagrams at
this particular kinematical point is necessary which is outside the
scope of the present considerations (for a general exposition see
\cite{BBT}). 

We can calculate the effect of this singularity in the following way.
For definiteness, let us specify $y>y^{'}$: then the integration
contour in (\ref{eq:euclideanpropagator}) can be closed in the $\Im m\,\rho<0$
half-plane giving the result\[
\,_{B}\left\langle 0\right|\Phi(x,y,\vec{r})\Phi(x^{'},y^{'},\vec{r}^{'})\left|0\right\rangle _{B}=-\int\frac{dk}{2\pi}\frac{Z}{2\sqrt{k^{2}+m^{2}}}\mathrm{e}^{-\sqrt{k^{2}+m^{2}}(y-y')}\left\{ -\frac{mg^{2}/2}{\sqrt{k^{2}+m^{2}}}\mathrm{e}^{-ik(x+x')}\right\} \]
where we have also performed the trivial $\vec{k}$ integration. Remembering
now that $x+x'<0$ we can close the contour in this term in the lower
half-plane which gives\[
\,_{B}\left\langle 0\right|\Phi(x,y,\vec{r})\Phi(x^{'},y^{'},\vec{r}^{'})\left|0\right\rangle _{B}\sim\frac{g^{2}}{8}Z\mathrm{e}^{m(x+x')}+\dots\]
and comparing to (\ref{eq:gtildecluster}) it follows that\[
\bar{g}=\frac{g}{2}\sqrt{\frac{Z}{2}}\]
 and so \[
\tilde{g}=\frac{g}{2}\]
It is useful to stress that for $D>1$ this result shows just the
consistency of the assumption (\ref{eq:2pt_sing_assumption}). In
$1+1$ dimensions, however, the exact reflection factors of many integrable
theories are known explicitely, and exhibit a pole at $\vartheta=i\pi/2$,
and in this case the cluster argument is in fact a field theoretic
proof of a relation between $g$ and $\tilde{g}$ (valid also for
the nonintegrable case), conjectured in \cite{dptw_onepoint} and
checked using both theoretical and numerical arguments in \cite{Lusch}.

\subsection{Finite size energy correction from the boundary state}

We now calculate the ground-state energy per transverse volume, $E_{0}^{\alpha\beta}(L)$,
of the system confined by two hyperplanes to the interval $0\leq x\leq L$
and subject to boundary conditions labelled by $\alpha$ and $\beta$
at the two ends. In doing so we compactify the other directions to
circles of perimeter $R$ with periodic boundary conditions and calculate
the partition function in two different ways, corresponding to the
two different Hamiltonian descriptions of the system introduced previously.
In the open channel (where $y=-it$ is the imaginary time) the partition
function can be written as \[
Z(L,R)=Tr(e^{-RH_{\alpha\beta}(L,R)})\]
which for large $R$ behaves as \[
\lim_{R\to\infty}Z(L,R)=e^{-E_{0}^{\alpha\beta}(L)V}+\textrm{small correction}\]
where $V=R^{D}$. In the closed channel (where $x=it$ is the Euclidean
time) the partition function is given by the following matrix element.
\[
Z(L,R)=\langle B_{\alpha}\vert e^{-LH(R)}\vert B_{\beta}\rangle\]
Inserting a complete set of eigenstates of the periodic (bulk) Hamiltonian
$H(R)$ \[
Z(L,R)=\sum_{n}\frac{\langle B_{\alpha}\vert n\rangle\langle n\vert B_{\beta}\rangle}{\langle n\vert n\rangle}e^{-E_{n}(R)L}\]
We concentrate on the leading finite size correction to the ground
state energy $E_{0}^{\alpha\beta}(L)$ and take $L$ to be large.
As a consequence the low lying energy levels dominate the sum:\[
Z(L,R)=e^{-E_{0}(R)L}\left[1+\sum_{\kappa,\vec{k}}\frac{\langle B_{\alpha}\vert\kappa,\vec{k}\rangle\langle\kappa,\vec{k}\vert B_{\beta}\rangle}{\langle\kappa,\vec{k}\vert\kappa,\vec{k}\rangle}e^{-\omega(\kappa,\vec{k})L}+\dots\right]\]
where the sum is over one-particle states with momentum $(\kappa,\vec{k}).$
The finite volume restricts the momentum to be $\kappa=\frac{2\pi}{R}n$,
and $k_{i}=\frac{2\pi}{R}n_{i}$ and the normalization of the creation
operators becomes \[
[A_{in}(\kappa,\vec{k}),A_{in}^{+}(\kappa^{'},\vec{k}^{'})]=V\omega(\kappa,\vec{k})\delta_{\kappa,\kappa^{'}}\delta_{\vec{k},\vec{k}^{'}}\]
 Using the form of the boundary state (\ref{eq:bstate_expanded})
yields \[
Z(L,R)=e^{-E_{0}(R)L}\left[1+mV\bar{K}_{\alpha}^{1}K_{\beta}^{1}e^{-mL}+\dots\right]\]
 We normalize the ground-state energy with periodic boundary condition
to zero: $E_{0}(R)=0$. As a result \emph{the ground state energy
per transverse volume} at leading order in $L$ has the form \begin{equation}
E_{0}^{\alpha\beta}(L)=-m\bar{K}_{\alpha}^{1}K_{\beta}^{1}e^{-mL}+\dots\label{Cas1}\end{equation}
If one of the $K^{1}$-s is zero then the leading correction comes
from two-particle states. Although we explained this situation in
\cite{Lusch,Casimir}, for completeness we recall the derivation of
the leading finite size correction. Using the cluster expansion (\ref{eq:bstate_expanded}),
the partition function can be written as\[
Z(L,R)=e^{-E_{0}(R)L}\left[1+\sum_{\kappa,\vec{k},\kappa^{'},\vec{k}^{'}}\frac{\langle B_{\alpha}\vert\kappa,\vec{k},\kappa^{'},\vec{k}^{'}\rangle\langle\kappa,\vec{k},\kappa^{'},\vec{k}^{'}\vert B_{\beta}\rangle}{\langle\kappa,\vec{k},\kappa^{'},\vec{k}^{'}\vert\kappa,\vec{k},\kappa^{'},\vec{k}^{'}\rangle}e^{-\omega(\kappa,\vec{k})L-\omega(\kappa^{'},\vec{k}^{'})L}+\dots\right]\]
The spectrum of the possible $\kappa,\kappa^{'}$ has to be determined
by solving the scattering problem of the two particles in volume $V$
exactly. However, in the infinite volume limit the interaction between
the particles can be neglected. (For large $L$ the main contribution
to the ground state energy comes from the low lying energy levels,
for which this approximation becomes exact as $R\rightarrow\infty$.)
Using the explicit form of the boundary state, the result for the
Casimir energy (per unit transverse area) is \[
E_{0}^{\alpha\beta}(L)=-\int\frac{d\vec{k}}{(2\pi)^{D-1}}\int_{0}^{\infty}\frac{d\kappa}{2\pi\omega(\kappa,\vec{k})}\bar{K}_{\alpha}^{2}(\kappa,\vec{k})K_{\beta}^{2}(\kappa,\vec{k})e^{-2\omega(\kappa,\vec{k})L}\]
as we explained in our previous paper \cite{Casimir}, where it was
derived using the assumption that the modes indexed by $\vec{k}$
can be treated as independent two-dimensional degrees of freedom (their
interaction only entering higher order terms explicitly). The present
derivation, however, completely dispenses with this additional assumption.

\section{Conclusion}

In this paper we have considered the relation between one-point function,
finite size corrections to ground state energy and the analytic structure
of scattering amplitudes in a general boundary QFT. Our main result
is the derivation {}``from first principles'' of the relations (\ref{eq:gbar_gtilde_intro},\ref{eq:gtilde_g_intro})
between the parameters characterizing these quantities for a general
$D+1$ dimensional quantum field theory.

Besides giving a solid theoretical background to preexisting results,
we developed the formalism of the boundary state for the $D+1$ dimensional
case, together with a new set of reduction formulae relating the boundary
state to the correlation functions, which can be used to express the
cluster expansion of the boundary state in terms of the boundary scattering
amplitudes.

The resulting formalism can be used to address various issues in boundary
quantum field theory. In addition to the derivation of the relation
between the $g$ parameters and its extension to general field theories,
we have shown that it can be used to derive the large volume asymptotics
of the Casimir effect. In particular in the case when there is no
one-particle coupling to the boundary, we proved that the result of
\cite{Casimir} for the leading behaviour is indeed universal, even
for interacting fields. We extended this result by giving the leading
term in the Casimir energy (\ref{Cas1}) for the case with nonvanishing
one-particle coupling, and used our results to relate it to the asymptotic
behaviour of the vacuum expectation value of the interpolating field.
It is important to note that it relates the planar Casimir effect
to physical quantities (vacuum expectation values, reflection factor)
that are calculable in the infinite volume boundary quantum field
theory, which is a much simpler setting than the finite volume case.
In addition, the expression is free of any ultraviolet divergences
right from the start, the underlying reason being that it involves
only relations between physical (i.e. renormalized) entities, as was
already discussed in \cite{Casimir}. It also noteworthy that it gives
a general formula for the dependence of the Casimir effect on the
material properties of the boundary (characterized by the reflection
factors) in the planar setting.

In principle, the cluster expansion for the boundary state (\ref{eq:bstate_expanded})
provides a systematic large volume expansion of the Casimir energy
(and other quantities, such as vacuum expectation values) if higher
particle terms are included. However, in the case when the reflection
factor has a singularity corresponding to the one-particle coupling,
even the two-particle term is divergent in general. This is a sort
of infrared divergence, which in the case of the ground state energy
is known to be eliminated by resummation of the series using e.g.
thermodynamic Bethe Ansatz (cf. \cite{Lusch} where a regular large
volume expansion is derived from boundary TBA), but this method only
works for integrable theories (or for theories in $D+1$ dimensions
with trivial bulk and completely elastic boundary scattering). It
is plausible that something similar happens in the case of nonintegrable
theories as well, but there is as yet no method to perform a resummation
of the relevant terms, which remains an interesting open problem.

\subsection*{Acknowledgements}

This research was partially supported by the EC network {}``EUCLID'',
contract number HPRN-CT-2002-00325, and Hungarian research funds OTKA
T043582, K60040 and TS044839. GT was also supported by a Bolyai J\'anos
research scholarship.

\appendix

\makeatletter 

\renewcommand\theequation{\hbox{\normalsize\Alph{section}.\arabic{equation}}} \@addtoreset{equation}{section}

\renewcommand\thefigure{\hbox{\normalsize\Alph{section}.\arabic{figure}}} \@addtoreset{figure}{section}

\renewcommand\thetable{\hbox{\normalsize\Alph{section}.\arabic{table}}} \@addtoreset{table}{section}

\makeatother

\section{Reduction formulae}

In this appendix we recall the derivation of the reduction formula
in the open channel and show its physical meaning, namely that it
connects the reflection factors to the on-shell part of the boundary
Green functions. Then we derive an analogous reduction formula in
the closed channel, and show that only the on-shell part of the boundary
Green function contributes to the boundary state, thus a relation
between the boundary state and the reflection factor is established.

\subsection{Reduction formula in the open channel}

The multi-particle reflection matrix is defined as the matrix element
\begin{eqnarray}
R_{k}^{l} & = & \,{}_{B}^{out}\langle p_{1},\vec{p}_{1};\dots;p_{k},\vec{p}_{k}\vert q_{1},\vec{q}_{1};\dots;q_{l},\vec{q}_{l}\rangle_{B}^{in}\nonumber \\
 & = & \,_{B}^{out}\langle p_{1},\vec{p}_{1};\dots;p_{k},\vec{p}_{k}\vert a_{in}^{+}(q_{1},\vec{q}_{1})\vert q_{2},\vec{q}_{2};\dots;q_{l},\vec{q}_{l}\rangle_{B}^{in}\label{matel}\end{eqnarray}
The creation operator of the initial particle state can be written
in terms of the asymptotic field as \[
a_{in}^{+}(k,\vec{k})=-\frac{i}{\sqrt{2}}\int_{-\infty}^{0}dx\cos(kx)\int d\vec{r}\, e^{i\vec{k}\vec{r}}\, e^{-i\omega(k,\vec{k})t}\overleftrightarrow{\partial_{t}}\Phi_{in}(x,t,\vec{r})\]
Using the property that the interpolating (interacting) field approaches
the free asymptotic fields as \[
\Phi(x,t,\vec{r})\to Z^{1/2}\Phi_{in/out}(x,t,\vec{r})\quad\mathrm{for}\quad t\to\mp\infty\]
 we obtain the following form of the reflection matrix \begin{eqnarray*}
R_{k}^{l} & = & _{B}^{out}\langle p_{1},\vec{p}_{1};\dots;p_{k},\vec{p}_{k}\vert a_{out}^{+}(q_{1},\vec{q}_{1})\vert q_{2},\vec{q}_{2};\dots;q_{l},\vec{q}_{l}\rangle_{B}^{in}+\\
 &  & i\sqrt{\frac{2}{Z}}\int_{-\infty}^{0}dx\int d\vec{r}e^{i\vec{q}_{1}\vec{r}}\int dt\partial_{t}\{\cos(q_{1}x)e^{-i\omega(q_{1},\vec{q}_{1})t}\overleftrightarrow{\partial_{t}}\,_{B}\langle out\vert\Phi(x,t,\vec{r})\vert in\rangle_{B}\}\end{eqnarray*}
where $_{B}\langle out\vert$, ($\vert in\rangle_{B}$) is the shorthand
form for $_{B}^{out}\langle p_{1},\vec{p}_{1};\dots;p_{k},\vec{p}_{k}\vert$
and $\vert q_{2},\vec{q}_{2};\dots;q_{l},\vec{q}_{l}\rangle_{B}^{in}$,
respectively. It is necessary to be careful when performing the partial
integration and keep the surface term. The connected part turns out
to be\begin{equation}
i\sqrt{\frac{2}{Z}}\int_{-\infty}^{0}dx\int dtd\vec{r}e^{-i\omega(q_{1},\vec{q}_{1})t}e^{i\vec{q}_{1}\vec{r}}\cos(q_{1}x)\{\partial_{t}^{2}-\partial_{x}^{2}-\vec{\partial^{2}}+m^{2}+\delta(x)\partial_{x}\}\,_{B}\langle out|\Phi(x,t,\vec{r})|in\rangle_{B}\label{connpart}\end{equation}
Repeating the same procedure one can express the reflection factor
as the product of integro-differential operators acting on the Green
functions. The connected part is related to the $N$-pont function,
while the disconnected one appears when at least one of the incoming
momentum (say $(q_{i},\vec{q}_{i})$) coincides with one of the outgoing
momentum (say $(p_{j},\vec{p}_{j})$). It contains a delta function
singularity $(2\pi)^{D}\omega(q_{i},\vec{q}_{i})\delta(q_{i}-p_{j})\delta(\vec{q}_{i}-\vec{p}_{j})$
whose coefficient is related to the $N-2$ point function. Here we
would like concentrate on the connected part and to show, that the
operator in (\ref{connpart}) truncates the leg of the momentum space
Green function and puts its momentum on-shell.

The momentum space Green function is defined by Fourier transformation\begin{eqnarray*}
\,_{B}\langle0\vert T_{t}\left(\Phi(x_{1},t_{1},\vec{r}_{1})\dots\Phi(x_{N},t_{N},\vec{r}_{N})\right)\vert0\rangle_{B} & =\\
 &  & \hspace{-5cm}\prod_{j=1}^{N}\int_{-\infty}^{\infty}\frac{d\omega_{j}}{2\pi}e^{i\omega_{j}t_{j}}\int_{-\infty}^{\infty}\frac{dk_{j}}{2\pi}e^{-ik_{j}x_{j}}\int_{-\infty}^{\infty}\frac{d\vec{k}_{j}}{(2\pi)^{D-1}}e^{-i\vec{k}_{j}\vec{r}_{j}}G_{bdry}^{N}(\{\omega_{i},k_{i},\vec{k}_{i}\})\end{eqnarray*}
Inspecting the perturbative expansion we can always write\begin{equation}
G_{bdry}^{N}(\{\omega_{i},k_{i},\vec{k}_{i}\})=(2\pi)^{D}\delta(\sum_{j}\omega_{j})\delta(\sum_{j}\vec{k}_{j})\left[\textrm{disc.}+\prod_{j}G_{bulk}^{2}(\omega_{j},k_{j},\vec{k}_{j})B_{bdry}^{N}(\{\omega_{i},k_{i},\vec{k}_{i}\})\right]\label{Greenf}\end{equation}
where the disconnected part contains at least one particle which is
unaffected by the boundary, that is a term proportional to $(2\pi)^{D}\delta(k_{i}-k_{j})\delta(\vec{k}_{i}-\vec{k}_{j})$.
\footnote{This definition of the disconnected part is slightly different from
the one used for the two-point case in our previous work \cite{BBT}
but is more convenient for multi-point functions.%
} Using the K\"allen-Lehman form of the bulk two-point function the
contribution of a single leg can be written as \begin{eqnarray}
\int\frac{d\omega}{2\pi}\frac{dk}{2\pi}\frac{d\vec{k}}{(2\pi)^{D-1}}e^{i\omega t}e^{-ikx}e^{-i\vec{k}\vec{r}}\label{1leg}\\
 & \left[\frac{iZ}{\omega^{2}-k^{2}-\vec{k}^{2}-m^{2}+i\epsilon}+\int_{2m}^{\infty}\frac{i\sigma(m^{'})dm^{'}}{\omega^{2}-k^{2}-\vec{k}^{2}-m^{'2}+i\epsilon}\right]B_{bdry}^{N}(\omega,k,\vec{k},\dots)\nonumber \end{eqnarray}
We are interested in the action of the operator \[
\int_{-\infty}^{0}dx\int dtd\vec{r}e^{-i\omega(q,\vec{q})t}e^{i\vec{q}\vec{r}}\cos(qx)\{\partial_{t}^{2}-\partial_{x}^{2}-\vec{\partial^{2}}+m^{2}+\delta(x)\partial_{x}\}\]
appearing in (\ref{connpart}) on (\ref{1leg}). Calculating the derivatives
$\partial_{t},\vec{\partial}$ and performing the $t,\vec{r}$ integrals
explicitely, we obtain $(2\pi)^{D}\delta(\omega(q,\vec{q})-\omega)\delta(\vec{q}-\vec{k})$,
and thus the $\omega$ and $\vec{k}$ integral replaces every $\omega$
by $\omega(q,\vec{q})$ and every $\vec{k}$ by $\vec{q}$. So it
remains to be shown that the operator \[
\int_{-\infty}^{0}dx\cos(qx)\left\{ -\partial_{x}^{2}-q^{2}+\delta(x)\partial_{x}\right\} \]
 operator truncates the leg \[
\int_{-\infty}^{\infty}\frac{dk}{2\pi}e^{-ikx}\left[\frac{iZ}{q^{2}-k^{2}+i\epsilon}+\int_{2m}^{\infty}\frac{i\sigma(m^{'})dm^{'}}{q^{'2}-k^{2}+i\epsilon}\right]B_{bdry}^{N}(\omega(q,\vec{q}),k,\vec{q},\dots)\]
where $q^{'2}=q^{2}+m^{2}-m^{'2}$. The actions of $-\partial_{x}^{2}-q^{2}$
and $\delta(x)\partial_{x}$ can be computed separately. The second
is simpler: it gives a factor $-ik$ and substitutes $x=0$ in the
$k$ integral. The term $-\partial_{x}^{2}-q^{2}$ gives a factor
$k^{2}-q^{2}$ but now the $x$ integral is nontrivial and we use
that the contour can be closed on the upper half plane giving\begin{eqnarray*}
2\int_{-\infty}^{0}dx\cos(qx)e^{-ikx+\epsilon x} & = & \frac{1}{-i(k-q)+\epsilon}+\frac{1}{-i(k+q)+\epsilon}\\
 & = & i\left(\mathcal{P}_{\frac{1}{k-q}}+\mathcal{P}_{\frac{1}{k+q}}\right)+\pi\left(\delta(k-q)+\delta(k+q)\right)\end{eqnarray*}
Consider first the terms containing the spectral density $\sigma$.
Due to the pre-factor $k^{2}-q^{2}$ the delta functions do not contribute
and the principal value can be replaced by the function itself, which
just kills the other $\sigma$ term coming from the $\delta(x)\partial_{x}$
term. Now we analyze the terms containing $Z$. In the formula for
$\delta(x)\partial_{x}$ we write\[
-ik\frac{iZ}{q^{2}-k^{2}+i\epsilon}=-\frac{Z}{2}\left[\frac{1}{k+q+i\epsilon}+\frac{1}{k-q-i\epsilon}\right]\]
The terms with the denominator $k+q+i\epsilon$ cancel, while the
ones with denominators $k-q\pm i\epsilon$ combine together to give
delta functions resulting in \[
-\frac{iZ}{2}B_{bdry}^{N}(\omega(q,\vec{q}),q,\vec{q},\dots)\]
which, combined with the pre-factor $i\sqrt{\frac{2}{Z}}$ in the
reduction formula, gives the contribution of one leg as \[
\sqrt{\frac{Z}{2}}B_{bdry}^{N}(\omega(q,\vec{q}),q,\vec{q},\dots)\]
Reduction of an outgoing particle gives rise to the same effect, so
collecting the contributions of all legs the connected part is simply
\begin{eqnarray}
R_{k}^{l} & = & (2\pi)^{D}\delta(\sum_{i}\omega(p_{i},\vec{p}_{i})-\sum_{j}\omega(q_{j},\vec{q}_{j}))\delta(\sum_{i}\vec{p}_{i}-\sum_{j}\vec{q}_{j})\left(\frac{Z}{2}\right)^{\frac{k+l}{2}}\times\nonumber \\
 &  & B_{bdry}^{N}(\omega(p_{i},\vec{p}_{i}),p_{i},\vec{p}_{i},\omega(q_{j},\vec{q}_{j}),q_{j},\vec{q}_{j})\label{Rkl}\end{eqnarray}
 In addition to the connected part treated above, there are two types
of disconnected parts: one from the reduction formula itself, and
the other originating from the disconnected part of the Green function
(\ref{Greenf}). Using induction in the number of particles it is
easy to show that the two contributions cancel each other, and therefore
(\ref{Rkl}) is the final answer to the reflection factor. It also
agrees with the results for the two-particle case in \cite{BBT} after
accounting for the difference in the normalization conventions.

\subsection{Reduction formula in the closed channel}

In the closed channel the Hilbert space is identical to that of the
bulk theory, and the quantities of interest are the multi-particle
matrix elements of the boundary state:\[
\langle B\vert q_{1},\vec{q}_{1};\dots;q_{N},\vec{q}_{N}\rangle_{in}=(2\pi)^{D}\delta(\sum_{j}q_{j})\delta(\sum_{j}\vec{q}_{j})\omega\left(q_{1},\vec{q}_{1}\right)K^{N}(\{ q_{i},\vec{q}_{i}\})\]
($\omega\left(q_{1},\vec{q}_{1}\right)$ is a normalization factor
introduced for convenience in order to conform with the conventions
for $K^{1}$ and $K^{2}$ in the main text). In the bulk theory matrix
elements are expressed in terms of the correlators via the reduction
formula, but due to the presence of a boundary in time they need to
be modified. First we express $A_{in}^{+}(q_{1},\vec{q}_{1})$ in
terms of the \emph{in} field as\[
\langle B\vert A_{in}^{+}(q_{1},\vec{q}_{1})\vert q_{2},\vec{q}_{2};\dots;q_{N},\vec{q}_{N}\rangle_{in}=-\frac{i}{\sqrt{2}}\int_{-\infty}^{\infty}dyd\vec{r}\, e^{-i\omega(q_{1},\vec{q}_{1})\tau+iq_{1}y+i\vec{q}_{1}\vec{r}}\overleftrightarrow{\partial_{\tau}}\langle B\vert\Phi_{in}(\tau,y,\vec{r})\vert in\rangle\]
where $\vert in\rangle$ is a shorthand for $\vert q_{2},\vec{q}_{2};\dots;q_{N},\vec{q}_{N}\rangle_{in}$.
Now we use that for large negative $\tau$ the interacting field approaches
the asymptotic field \[
\lim_{\tau\to-\infty}\Phi_{in}(\tau,y,\vec{r})=Z^{-1/2}\Phi(\tau,y,\vec{r})\]
 and also\[
f(\tau)=f(0)-\int_{\tau}^{0}\partial_{\tau}fd\tau^{'}\]
 to obtain the rule for the elimination of a particle from the initial
state: \begin{eqnarray*}
\langle B\vert A_{in}^{+}(q_{1},\vec{q}_{1})\vert q_{2},\vec{q}_{2};\dots;q_{N},\vec{q}_{N}\rangle_{in} & = & \frac{i}{\sqrt{2Z}}\int_{-\infty}^{\infty}dy\int_{-\infty}^{0}dt\int d\vec{r}\, e^{-i\omega(q_{1},\vec{q}_{1})\tau+iq_{1}y+i\vec{q}_{1}\vec{r}}\\
 &  & \hspace{-2cm}\left\{ \partial_{\tau}^{2}-\partial_{y}^{2}-\vec{\partial}^{2}+m^{2}-\delta(\tau)(\partial_{\tau}+i\omega(q_{1},\vec{q}_{1}))\right\} \,\langle B\vert\Phi(\tau,y,\vec{r})\vert in\rangle\end{eqnarray*}
Applying this procedure successively, the multi-particle matrix element
can be expressed as the product of integro-differential operators
\[
\frac{i}{\sqrt{2Z}}\int_{-\infty}^{\infty}dy\int_{-\infty}^{0}dt\int d\vec{r}\, e^{-i\omega(q,\vec{q})\tau+iqy+i\vec{q}\vec{r}}\left\{ \partial_{\tau}^{2}-\partial_{y}^{2}-\vec{\partial}^{2}+m^{2}-\delta(\tau)(\partial_{\tau}+i\omega(q,\vec{q}))\right\} \]
 acting on the correlator. The reduction formula obtained this way
differs from its bulk counterpart by the presence of the $\delta(\tau)$
term. It also differs from the analogous expression (\ref{connpart})
in the open channel by containing $e^{-i\omega(q,\vec{q})\tau}$ instead
of $\cos(kx)$ and by the extra $-i\delta(\tau)\omega(q,\vec{q})$
term. Despite these differences it can be shown that it also truncates
the momentum space Green function and puts the momentum on-shell.
To do so we rewrite the open channel one leg contribution to the Green
function (\ref{1leg}) in terms of the closed channel \begin{eqnarray*}
\int_{-\infty}^{\infty}\frac{d\omega}{2\pi}e^{i\omega\tau}\frac{dk}{2\pi}e^{-ikx}\frac{d\vec{k}}{(2\pi)^{D-1}}e^{-i\vec{k}\vec{r}}\\
 & \hspace{-2cm}\left[\frac{iZ}{\omega^{2}-k^{2}-\vec{k}^{2}-m^{2}+i\epsilon}+\int_{2m}^{\infty}\frac{i\sigma(m^{'})dm^{'}}{\omega^{2}-k^{2}-\vec{k}^{2}-m^{'2}+i\epsilon}\right]B_{bdry}^{N}(-ik,i\omega,\vec{k},\dots)\end{eqnarray*}
In parallel with the open channel calculation we can eliminate the
dependence on $k$, and $\vec{k}$ by performing explicitly the differentiations
and integrations. Finally it remains to show that the operator \[
\int_{-\infty}^{0}d\tau\,\, e^{-i\omega(q,\vec{q})\tau}\left\{ \partial_{\tau}^{2}+\omega(q,\vec{q})^{2}-\delta(t)\left(i\omega(q,\vec{q})+\partial_{\tau}\right)\right\} \]
truncates the leg \[
\int_{-\infty}^{\infty}\frac{d\omega}{2\pi}e^{i\omega\tau}\left[\frac{iZ}{\omega^{2}-\omega(q,\vec{q})^{2}+i\epsilon}+\int_{2m}^{\infty}\frac{i\sigma(m^{'})dm^{'}}{\omega^{2}-\omega(q^{'},\vec{q})^{2}+i\epsilon}\right]B_{bdry}^{N}(-iq,i\omega,\vec{q},\dots)\]
where, as before, $q^{'2}=q^{2}+m^{2}-m^{'2}$. We compute separately
the contributions from $\partial_{\tau}^{2}+\omega(q,\vec{q})^{2}$
and $-\delta(\tau)\left(i\omega(q,\vec{q})+\partial_{\tau}\right)$.
The second is simpler: it gives a factor $-i(\omega(q,\vec{q})+\omega)$
and substitutes $\tau=0$ in the $\omega$ integration. The operator
$\partial_{\tau}^{2}+\omega(q,\vec{q})^{2}$ gives the factor $(-\omega^{2}+\omega(q,\vec{q})^{2})$
and the $\tau$ integration gives \[
\int_{-\infty}^{0}d\tau\,\, e^{-i\omega(q,\vec{q})\tau}e^{i\omega\tau+\epsilon\tau}=\frac{-i}{\omega-\omega(q,\vec{q})-i\epsilon}=-i\mathcal{P}_{\frac{1}{\omega-\omega(q,\vec{q})}}+\pi\delta(\omega-\omega(q,\vec{q}))\]
 Let us first concentrate on the $\sigma$ terms. The $\delta$ function
does not contribute due to the pre-factor $-\omega^{2}+\omega(q,\vec{q})^{2}$,
while the principal value can be replaced with the function itself
and this just kills the $\sigma$ term coming from the $\delta(\tau)$
part. In the $Z$ term we use that \[
-i\frac{\omega+\omega(q,\vec{q})}{\omega^{2}-\omega(q,\vec{q})^{2}+i\epsilon}=\frac{-i}{\omega-\omega(q,\vec{q})+i\epsilon}\]
This kills the principal value term above and results in a factor
two in front of the $\delta\left(\tau\right)$ part, which together
gives \[
-iZB_{bdry}^{N}(-iq,i\omega(q,\vec{q}),\vec{q},\dots)\]
Combining with the pre-factor $\frac{i}{\sqrt{2Z}}$ in the reduction
formula gives the contribution of one leg as \[
\sqrt{\frac{Z}{2}}B_{bdry}^{N}(-iq,i\omega(q,\vec{q}),\vec{q},\dots)\]
 In the closed channel we have just one type of disconnected parts,
the one coming from the Green functions (\ref{Greenf}), which contains
at least one bulk two-point function as a disconnected piece. Just
as in the two-particle case this does not contribute via the reduction
formula, and since it appears multiplicatively the whole contribution
coming from the disconnected part is vanishing. As a consequence only
the connected part of the Green function contributes to the boundary
state, which is therefore identical to the $q\leftrightarrow i\omega$
continuation of the result (\ref{Rkl}) for the reflection factor,
and so $K^{k+l}$ can be expressed in terms of $R_{l}^{k}$. We make
this relation explicit for the two-particle term $K^{2}$ in subsection
3.3.

\end{document}